\documentclass[12pt]{article}
\pdfoutput=1

\usepackage[height=23truecm,width=16.4truecm,top=2.2truecm]{geometry}

\usepackage[utf8]{inputenc}
\usepackage{graphicx}
\usepackage{amsmath}
\usepackage{amssymb}
\usepackage{hyperref}
\hypersetup{colorlinks,citecolor=TokiwaIro,linkcolor=RuriIro,urlcolor=RuriIro,bookmarksopen,bookmarksnumbered,pdfstartview=FitH,linktocpage}
\usepackage{cite}
\usepackage{braket}
\usepackage{bm}
\usepackage{bbm}
\usepackage{color}
\usepackage[svgnames,psnames,table]{xcolor}
\usepackage{float}
\usepackage{hhline}
\usepackage{multirow}
\usepackage{indentfirst}
\usepackage{latexsym}
\usepackage{mathtools}
\usepackage{cancel}
\usepackage{slashed}
\usepackage{colortbl}
\usepackage[mathscr]{euscript}
\usepackage{enumitem}
\usepackage{subcaption}
\usepackage{comment}
\usepackage{amsthm}

\usepackage{ulem}

\numberwithin{equation}{section}

\allowdisplaybreaks

\definecolor{RuriIro}{rgb}{0.,0.28,0.60}
\definecolor{TokiwaIro}{rgb}{0.,0.39,0.16}

\definecolor{dred}{rgb}{0.7,0.15,0.09}
\definecolor{dblue}{rgb}{0,0.12,0.64}
\definecolor{dgreen}{rgb}{0.2,0.51,0.19}
\definecolor{winered}{rgb}{0.58,0.18,0.27}

\newcommand{\nn}{\nonumber}

\newcommand{\mc}{\mathcal}
\newcommand{\mr}{\mathrm}

\newcommand{\mb}{\mathbf}
\newcommand{\mbb}{\mathbb}
\newcommand{\mbbm}{\mathbbm}
\newcommand{\msc}{\mathscr}

\newcommand{\del}{\partial}
\newcommand{\ol}{\overline}
\newcommand{\dd}{\mathrm{d}}
\newcommand{\ee}{\mathrm{e}}
\newcommand{\iu}{\mathrm{i}}

\newcommand{\ttr}{\mathrm{t}}

\newcommand{\moduli}{\mathrm{moduli}}
\newcommand{\SL}{\mathrm{SL}}

\newcommand{\NL}{\mathrm{NL}}

\begin{document}

\begin{titlepage}

\begin{flushright}
EPHOU-23-005
\\
KYUSHU-HET-256
\end{flushright}

\vspace{1cm}

\begin{center}

{\Large \bfseries
Moduli inflation from modular flavor symmetries
}

\vspace{1cm}

\renewcommand{\thefootnote}{\fnsymbol{footnote}}
{%
\hypersetup{linkcolor=black}
Yoshihiko Abe$^{1}$\footnote[1]{yabe3@wisc.edu},
\ 
Tetsutaro Higaki$^{2}$\footnote[2]{thigaki@rk.phys.keio.ac.jp},
\ 
Fumiya Kaneko$^{2}$\footnote[3]{f7m1y9@keio.jp}
\\
Tatsuo Kobayashi$^{3}$\footnote[4]{kobayashi@particle.sci.hokudai.ac.jp},
\ 
and Hajime Otsuka$^{4}$\footnote[5]{otsuka.hajime@phys.kyushu-u.ac.jp}
}%
\vspace{8mm}

{\itshape
$^1${Department of Physics, University of Wisconsin-Madison, Madison, WI 53706, USA}\\
$^2${Department of Physics, Keio University, Yokohama 223-8533, Japan}\\
$^3${Department of Physics, Hokkaido University, Sapporo 060-0810, Japan}\\
$^4${Department of Physics, Kyushu University, 744 Motooka, Nishi-ku, Fukuoka 819-0395, Japan}
}

\vspace{8mm}

\end{center}

\abstract{
We study slow-roll inflation model controlled by the modular flavor symmetry. 
In the model, the modulus field plays a role of inflaton
and the introduction of the stabilizer field
coupled to a modular form in the superpotential
produces the inflaton potential.
In order to generate the flat direction for the slow-roll inflation, we consider the K\"{a}hler potential corrected by the modular form. 
It is noted that the modulus field perpendicular to the inflaton direction is stabilized during the inflation.
The model turns out to be consistent with the current observations and behaves similarly to the $\alpha$-attractor models in some parameter spaces. 
The inflaton rolls down to the CP-symmetric vacuum at the end of inflation.
}

\end{titlepage}

\renewcommand{\thefootnote}{\arabic{footnote}}
\setcounter{footnote}{0}
\setcounter{page}{1}

\tableofcontents

\section{Introduction}

The origin of the flavor structure of quarks and leptons is one of the big mysteries in particle physics.
Recently, the flavor symmetry based on the modular group~\cite{Feruglio:2017spp} attracts much attention.
In these models, the three generations of quarks and leptons transform non-trivially under the modular symmetry, that is, the modular symmetry is in a sense a flavor symmetry.
On top of that, Yukawa couplings are assumed to be modular forms, which are holomorphic functions of the modulus $\tau$ and non-trivially transform under the action of the modular group.
As discussed in Ref.~\cite{deAdelhartToorop:2011re}, it is remarkable that the (in)homogeneous finite modular group $\Gamma_N^{(\prime)}$ with the level $N \leq 5$ is isomorphic to the well-known (double-covering of) permutation group, such as $S_3$, $A_4^{(\prime)}$, $S_4^{(\prime)}$, and $A_5^{(\prime)}$, which have been intensively studied to explain the lepton flavor structure in the literature~\cite{Feruglio:2017spp,Kobayashi:2018vbk,Penedo:2018nmg,Novichkov:2018nkm,Ding:2019xna,Liu:2019khw,Novichkov:2020eep,Liu:2020akv,Liu:2020msy}.
These non-Abelian finite groups have been studied in flavor models for quarks and leptons~\cite{Altarelli:2010gt,Ishimori:2010au,Ishimori:2012zz,Hernandez:2012ra,King:2013eh,King:2014nza,Tanimoto:2015nfa,King:2017guk,Petcov:2017ggy,Feruglio:2019ybq,Kobayashi:2022moq}.

\medskip
The modular symmetry is well-motivated from the higher dimensional theories such as superstring theory.
For example, if we consider the torus or its orbifold compactification, the modulus parameter $\tau$ is the complex structure modulus, which is a dynamical degree of freedom of the effective field theory determining the shape of the torus.
The modular symmetry appears as the geometrical symmetry associated with this compact space. 
The Yukawa couplings are obtained by the overlap integral of the profile functions of the matter zero-modes and expressed as the function of the modulus which transforms non-trivially under the modular transformation.
Hence, the vacuum expectation value of the modulus determines the flavor structure and therefore should be stabilized.
The behavior of the zero-mode function under the modular transformation was studied in magnetized D-brane models~\cite{Kobayashi:2017dyu,Kobayashi:2018rad,Kobayashi:2018bff,Ohki:2020bpo,Kikuchi:2020frp,Kikuchi:2020nxn,Hoshiya:2020hki}, heterotic orbifold models~\cite{Lauer:1989ax,Lauer:1990tm,Ferrara:1989qb,Baur:2019kwi,Nilles:2020nnc,Nilles:2020gvu} and heterotic string theory on Calabi-Yau threefolds \cite{Ishiguro:2020nuf,Ishiguro:2021ccl}.
The modular flavor symmetric three-generation models based on the magnetized extra dimension were discussed in Refs.~\cite{Hoshiya:2020hki,Kikuchi:2021ogn}.
The modulus stabilization is also discussed in Refs.~\cite{Ishiguro:2020tmo,Novichkov:2022wvg}.

\medskip
The modulus field can be a candidate of the inflaton to realize the inflationary expansion of the early Universe,
because modular symmetry acting on it includes a shift symmetry which tends to flatten its scalar potential. 
In this work, we consider the inflation model controlled by the modular flavor symmetry.\footnote{The hybrid inflation induced by one of right-handed sneutrinos was discussed in the context of modular flavor symmetry \cite{Gunji:2022xig}.}
The modulus field plays the role of inflaton and its profile is given as the trajectory in the complex plane associated with the complex modulus.
Indeed, the inflation driven by modulus field was studied in modular symmetric supergravity model \cite{Kobayashi:2016mzg}.
Also, non-supersymmetric models using modular forms were studied in~\cite{Higaki:2015kta,Schimmrigk:2016bde,Lynker:2019joa,Schimmrigk:2021tlv}.\footnote{
See also, e.g., \cite{Abe:2014xja} for a supersymmetric model.
}
In this paper, the stabilizer field is introduced in order to generate the scalar potential, which is assumed to be the singlet representation of $\Gamma_N$ but has the non-trivial weight.
We find that the K\"{a}hler potential corrected by the modular form makes the scalar potential flatter and realizes slow-roll inflation which is consistent with the current observations.\footnote{
The realization of the de Sitter spacetime in string theory is difficult to achieve \cite{Dine:1985he,Obied:2018sgi,Ooguri:2018wrx}, and there is still room for discussion (see also Refs.~\cite{Palti:2019pca,vanBeest:2021lhn,Agmon:2022thq} for the reviews.)
In the context of the modular symmetry, see Ref.~\cite{Leedom:2022zdm}.
}
Then, the modulus field perpendicular to the inflaton direction is stabilized during the inflation, evading the overshooting problem \cite{Brustein:1992nk}.
Furthermore, the slow-roll $\alpha$-attractor solution 
\cite{Kallosh:2013hoa,Kallosh:2013yoa,Galante:2014ifa,Kallosh:2014rga,Kallosh:2015lwa,Linde:2015uga,Carrasco:2015pla}
is realized in some parameter spaces. 
At the end of the inflation, the modulus turns out to be stabilized at the CP-conserving vacuum. This can be favored in terms of the flavor structure~\cite{Okada:2019uoy} and regarded as a generic consequence in a modular invariant scalar potential \cite{Font:1990nt,Ferrara:1990ei,Cvetic:1991qm,Gonzalo:2018guu}.

\medskip
The rest part of this paper is organized as follows.
In Sec.~\ref{sec:modular-symmetry}, we give a brief review of the modular flavor symmetry.
In Sec.~\ref{sec:model}, we introduce inflation model based on the modular flavor symmetry.
In Sec.~\ref{sec:inflation}, 
we discuss the correction of the modular form in the K\"{a}hler potential and the deformation of the potential via this correction.
We study the inflationary dynamics of the modulus field and show the parameter space of our model.
Sec.~\ref{sec:conclusion} is devoted to our conclusions.
In App.~\ref{app:modular-forms}, we exhibit modular forms of the finite modular group $\Gamma_N$ with $N =3,4,5$. 
In App.~\ref{app:multi-field-inflation},
the formulae of the multi-field inflation are summarized.
In App.~\ref{app:Y^{3,6}-and-Y^{3,4}},
a model with two modular forms in the superpotential is discussed.
In App.~\ref{app:other-vac-inflation}, we show the inflationary dynamics rolling into the vacuum, which is identical to the vacuum discussed in Secs.~\ref{sec:model} and \ref{sec:inflation} via the modular transformation.

\section{Modular flavor symmetry}
\label{sec:modular-symmetry}

In this section, we give a brief review of the modular flavor symmetry.
The homogeneous modular group $\Gamma \coloneqq \SL(2,\mbb{Z})$ is defined by 
\begin{align}
	\Gamma \coloneqq \left\{
		\begin{pmatrix}
			a & b \\
			c & d
		\end{pmatrix}
		\bigg|
		\ 
		a, b, c, d \in \mbb{Z},
		\quad 
		ad -bc = 1
	\right\}.
\end{align}
This is generated by 
\begin{align}
	S = \begin{pmatrix}
		0 & 1 \\
		-1 & 0
	\end{pmatrix},
	\qquad
	T = \begin{pmatrix}
		1 & 1 \\
		0 & 1
	\end{pmatrix},
	\quad 
	R = \begin{pmatrix}
		-1 & 0 \\
		0 & -1
	\end{pmatrix},
\end{align}
which satisfy the following relations:
\begin{align}
	S^2 = R,
	\quad
	(ST)^3 = R^2 = S^4 = \mbbm{1},
	\quad
	TR =RT.
\end{align}
Under the $\SL(2,\mbb{Z})$ transformation, the modulus $\tau$ transforms as 
\begin{align}
	\gamma: \tau \mapsto \gamma \tau = \frac{a \tau + b}{c \tau + d},
	\qquad 
	\gamma \in \SL(2,\mbb{Z}).
	\label{eq:modular-transformation-tau}
\end{align}
In particular, the action of the generators of $\Gamma$, $S$ and $T$, are written as 
\begin{align}
	S: \tau \mapsto - \frac{1}{\tau},
	\qquad 
	T: \tau \mapsto \tau + 1,
	\label{eq:S-T-transformation}
\end{align}
and $\tau$ is invariant under $R=-\mbbm{1}$.
This transformation is called the modular transformation and $\bar{\Gamma} \coloneqq \Gamma /\mbb{Z}_2^R$ is called modular group, where $\mbb{Z}_2^R$ is generated by $R$.
It is noted that $T$ is similar to a shift symmetry often discussed in axion models \cite{DiLuzio:2020wdo}.

\medskip
The congruence subgroup of the level $N$, denoted by $\Gamma(N)$, is defined by 
\begin{align}
	\Gamma(N) \coloneqq \left\{
		\begin{pmatrix}
			a & b \\
			c & d
		\end{pmatrix} \in \SL(2,\mbb{Z}),
		\quad
		\begin{pmatrix}
			a & b \\
			c & d
		\end{pmatrix}
		\equiv 
		\begin{pmatrix}
			1 & 0 \\
			0 & 1
		\end{pmatrix}
		\mod N
	\right\}.
\end{align}
The quotients $\Gamma_N \coloneqq \bar{\Gamma}/ \Gamma(N)$ for $N=2,\ 3,\ 4$, and 5 are respectively isomorphic to $S_3$, $A_4$, $S_4$, and $A_5$~\cite{deAdelhartToorop:2011re}.
In addition, the quotients $\Gamma'_N \coloneqq \Gamma / \Gamma(N)$ for $N=3,4,$ and 5 are isomorphic to $A'_4$, $S'_4$, and $A'_5$, which are double covering groups of $A_4$, $S_4$, and $A_5$.
In these quotients, $T$ satisfies $T^N = \mbbm{1}$, which generates $\mbb{Z}_N^T$ symmetry.

\medskip
Hereafter, we focus on the supergravity (SUGRA) formulation \cite{Wess:1992cp,Ferrara:1989bc} for concreteness.
Under the modular transformation, a matter (super)field $\Phi$
with the modular weight $k_\Phi$ transforms as 
\begin{align}
    \Phi \mapsto  (c \tau + d)^{k_\Phi} \rho(\gamma) \Phi,
    \label{eq:modular-transformation-Phi}
\end{align}
where $\rho(\gamma)$ denotes the representation matrix determined by the representation of $\Phi$ for $\Gamma(N)$.
Here and hereafter, we use the convention that the superfield and its lowest component are 
denoted by the same letter.
A modular form $Y(\tau)$, which depends on $\tau$, similarly transforms under the modular flavor symmetries.
The matter K\"{a}hler potential is assumed to be given by 
\begin{align}
	\mc{K}_\Phi = \frac{|\Phi|^2}{(-\iu \tau + \iu \bar{\tau})^{-k_\Phi}},
\end{align}
which is invariant under the transformations \eqref{eq:modular-transformation-tau} and \eqref{eq:modular-transformation-Phi}.
Later, we will consider the correction which is dependent of the modular form.

\medskip
The K\"{a}hler potential for the modulus field
typically has the following form,
\begin{align}
	\mc{K}_\moduli = - M_P^2 h \log (- \iu \tau + \iu \bar{\tau}),
	\label{eq:Kahler-potential-tau}
\end{align}
and $M_P$ denotes the reduced Planck scale, $M_P \approx 2.4 \times 10^{18}~\mr{GeV}$.
In the following parts, we will set $M_P= 1$ otherwise stated.
Here, $h$ is a dimensionless constant, which is related to the choice of the extra dimension in the higher dimensional theory.
In the toroidal compactification, for example, it is found that $h=1$ for the complex structure or K\"{a}hler modulus on $\mbb{T}^2$ and $h=3$ for the overall complex structure or K\"{a}hler modulus on $\mbb{T}^6$.\footnote{
The modulus field $\tau$ in this work is assumed to be general complex modulus which parameterizes the size or the shape of the torus.
In heterotic string theory, both complex structure moduli and K\"{a}hler moduli play this role.
In type II theory, the shape and volume moduli play that in type IIB and IIA, respectively.
If there are multiple moduli fields, the K\"{a}hler potential is given by the following form:
\begin{align}
    \mc{K}_\moduli = -M_P^2 \sum_i h_i \log (- \iu \tau_i + \iu \bar{\tau}_i ),
\end{align}
where $i$ denotes the label of the moduli.
In this paper, we focus on the dynamics of the single complex modulus and the K\"{a}hler potential is assumed to be given by Eq.~\eqref{eq:Kahler-potential-tau}.
}
Then, the kinetic term is given by
\begin{align}
	\bm{L}_{\moduli,\mr{kin.}} 
	= - \frac{h}{4 \tau_I^2} \del_\mu \tau \del_\nu \bar{\tau} g^{\mu\nu} 
    = - \frac{h}{4 \tau_I^2} \bigl[
		(\del_\mu \tau_R)^2 + (\del_\mu \tau_I)^2
	\bigr],
    \label{eq:kinetic-term-modulus}
\end{align}
where we decompose the modulus as $\tau = \tau_R + \iu \tau_I$.
Under the modular transformation \eqref{eq:modular-transformation-tau}, this K\"{a}hler potential transforms as
\begin{align}
	\mc{K}_\moduli \mapsto \mc{K}_\moduli + h f + h \bar{f},
	\qquad
	f = \log (c \tau + d).
\end{align}
The invariance under this K\"{a}hler transformation requires that the superpotential $W$ should transform as 
\begin{align}
	W \mapsto \ee^{-hf} W = (c \tau + d)^{-h} W.
\end{align}
We find that the superpotential is the modular form with the weight $- h$ from this equation.

\section{Model}
\label{sec:model}

\subsection{Scalar potential}

In this section, we introduce the inflation model 
controlled by the modular flavor symmetry.
As the inflaton field we focus on a complex structure modulus associated with this modular symmetry, since
the modular symmetry includes the $T$ transformation 
$\tau \mapsto \tau +1$, which tends to flatten the scalar potential and makes it suitable for the slow-roll inflation.
The total bosonic action is given by 
\begin{align}
	S = \int \dd^4x \sqrt{-g} \biggl[
        \frac{1}{2} \mc{R}
        - \frac{1}{2} K_{ab} (\phi) \del_\mu \phi^a \del_\nu \phi^b g^{\mu\nu} - V(\phi)
	\biggr],
\end{align}
where $\mc{R}$ is the Ricci scalar.
$K_{ab}$ denotes the scalar field metric of this modulus field, and $\phi^a = ( \tau_R, \tau_I)^\ttr$.
In this work, $K_{ab}$ is derived from Eq.~\eqref{eq:Kahler-potential-tau} and has the following form:
\begin{align}
    K_{ab} = \frac{h}{2 \tau_I^2} \begin{pmatrix}
        1 & 0 \\
        0 & 1
    \end{pmatrix},
    \qquad
    K^{ab} = (K_{ab})^{-1} = \frac{2 \tau_I^2}{h} \begin{pmatrix}
    		1 & 0 \\
    		0 & 1
    \end{pmatrix}.
	\label{eq:Kahler-metric-tau}
\end{align}
The scalar potential is given by
\begin{align}
    V = \ee^{\mc{K}} \Bigl[
        \mc{K}^{\alpha \bar{\beta}} D_\alpha W \ol{D_\beta W} - 3 |W|^2 
    \Bigr],
\end{align}
where $\mc{K}_{\alpha \bar{\beta}}$ is the total K\"{a}hler metric.
$D_I$ acts on the superpotential as $D_\alpha W = \del_\alpha W + W \del_\alpha \mc{K}$ with the K\"{a}hler potential $\mc{K}$.
The indices $\alpha, \beta$ run all the superfield components, $\alpha, \beta = \tau, X$.
In order to generate the scalar potential for the modulus field, we introduce a matter field, so-called a stabilizer field, which is a trivial singlet but has a non-trivial weight under the modular transformation.
We consider the following superpotential proportional to  the stabilizer field $X$~\cite{Kallosh:2010ug,Kallosh:2010xz,Kallosh:2014vja,Ferrara:2014kva,Higaki:2016ydn},\footnote{
See also Refs.~\cite{Izawa:1996pk,Intriligator:1996pu,Intriligator:2006dd,Kitano:2006wz} for the supersymmetry breaking field.
}
\begin{align}
	W = \Lambda^2 Y(\tau) X.
	\label{eq:superpotential-form}
\end{align}
Here, $Y(\tau)$ is a modular form which is a trivial singlet under the modular transformation $\Gamma_N$ and has a non-zero weight, and $\Lambda$ is a constant of an energy scale characterizing this interaction.
As discussed in the previous section, the modular weights of $Y(\tau)$ and $X$, denoted by $k_Y$ and $k_X$, respectively, satisfy $-h = k_Y + k_X$ due to the invariance under the K\"{a}hler (modular) transformation.
This model is applicable to a general $\Gamma_N$ flavor symmetry because the dimension of the trivial singlet modular forms is unity up to the weight 10.
We summarize the modular forms of level 3 ($A_4$), 4 ($S_4$), and 5 ($A_5$) cases as the reference in App.~\ref{app:modular-forms}.
You can find that the trivial singlets share the same form
and can be given by Eisenstein functions.
For the number of the trivial singlets with the higher weight, see Ref.~\cite{Shimura:1971mod}.
In the following discussion, we use the singlet with the weight 
\begin{align}
k_Y=6, 
\end{align}
denoted by $Y^{3,6}_{\bm{1}}$~\cite{Feruglio:2017spp} as the concrete modular form. 
In this case, the modulus field tends to have a minimum around $\tau \sim \iu$ at $|Y^{3,6}_{\bm{1}}| \ll 1$ as explicitly shown in the next section, which can be favored from the flavor structure~\cite{Okada:2019uoy} as well as the moduli stabilization \cite{Ishiguro:2020tmo}.
Furthermore, the stabilizer field $X$ determines the magnitude of supersymmetry (SUSY) breaking as discussed below.

\medskip
The weight of the stabilizer field is given by $k_X = - h -6$ for a given $h$.
The scalar potential becomes
\begin{align}
	V \approx \Lambda^4 \ee^{\mc{K}} \mc{K}_{X \bar{X}}^{-1} |Y^{3,6}_{\bm{1}}(\tau)|^2
	= \Lambda^4 (2 \tau_I)^{k_Y} |Y^{3,6}_{\bm{1}}(\tau)|^2 = \Lambda^4 ( 2 \tau_I)^6 |Y^{3,6}_{\bm{1}}(\tau)|^2,
	\label{eq:model0-potential}
\end{align}
where we assumed $X \ll 1$ during the inflation and the K\"{a}hler potential is 
\begin{align}
	\mc{K} = - h \log ( -\iu \tau + \iu \bar{\tau}) + \frac{|X|^2}{(-\iu \tau + \iu \bar{\tau})^{-k_X}}.
	\label{eq:model0-Kahlerpotential}
\end{align}
We note that the overall modulus dependence in the scalar potential coming from the K\"{a}hler potential and K\"{a}hler metric are determined by the weight of $Y$ due to the relation $-h = k_Y + k_X$. 
Unlike the ordinary moduli potential arisen from the dimensional compactification, 
this potential does not have the runaway structure in $\tau_I \to \infty$ limit.\footnote{
The no runaway behavior in the modular invariant scalar potential is discussed in Ref.~\cite{Carrasco:2015pla}.
See also \cite{Gonzalo:2018guu} for a recent discussion.
}
In this scalar potential, the explicit values of $h$ and $k_X$ are not relevant, but $h$ determines the normalization of the modulus in their kinetic term.
Hereafter throughout this paper, we choose 
\begin{align}
h = 2 \qquad \to \qquad k_X = -8.
\end{align}

\begin{figure}[t]
	\centering
	\includegraphics[width=0.48\textwidth]{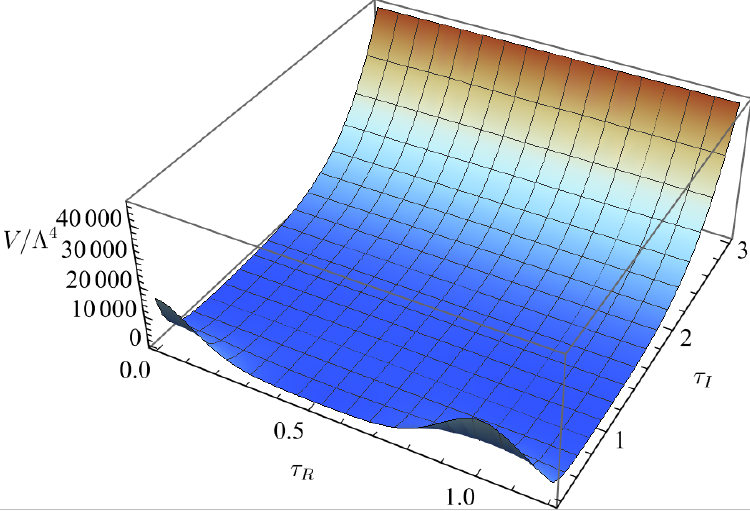}
	\caption{The 3D plot of the scalar potential \eqref{eq:model0-potential}.}
	\label{fig:model0-potential}
\end{figure}

\medskip
The 3D plot of the potential \eqref{eq:model0-potential} is shown in Fig.~\ref{fig:model0-potential}.
It is found that $\tau_R$ feels the axion cosine potential.
In the $\tau_I$ direction, this potential has the exponential dependence in addition to the $(2 \tau_I)^{k_Y}$ overall contribution.
This behavior can be easily seen from the $q$ expansion of the modular form.
The modular form is a function of $q \coloneqq \ee^{2 \pi \iu \tau}$.
In the $|q| \ll 1$ or $\tau_I \gg 1$ region, we find the following approximate form\footnote{
Throughout this paper, we have used $q$-expansion 
up to ${\cal O}(q^{12})$ in the modular forms and the scalar potential, since we have not find a change in the numerical calculation when comparing such expansions with those of ${\cal O}(q^{n})$ with $n< 12$.
}
\begin{align}
	|Y^{3,6}_{\bm{1}}(q)|^2 &= |1 - 504 q - 16632 q^2 + \mc{O}(q^3)|^2 
	\nn \\
	&= 1 + 504^2 \ee^{-4 \pi \tau_I} + 16632^2 \ee^{- 8 \pi \tau_I}
	\nn \\
	& \qquad 
	+ 2 (-504 \ee^{-2 \pi \tau_I} + 504 \times 16632 \ee^{- 6 \pi \tau_I} ) \cos (2 \pi \tau_R)
	\nn \\
	& \qquad 
	+ 33264 \ee^{-4 \pi \tau_I} \cos (4 \pi \tau_R) + \cdots.
	\label{eq:q-expansion-Y^{3,6}_1}
\end{align}
The equation of
$V \sim |Y^{3,6}_{\bm{1}}|^2 \sim |1 - 504 q|^2 \sim 0$
shows that the modulus field tends to be stabilized around the $\tau = \iu$, 
which is the CP-conserving vacuum.\footnote{
If $Y^{3,4}_{\bm{1}}$ is used instead of $Y^{3,6}_{\bm{1}}$,
the potential minimum will be given by 
$Y^{3,4}_{\bm{1}} \sim 1 + 240 q \sim 0$ 
and the CP-conserving vacuum will be realized around 
$\tau \sim 1/2 + \iu$.
}

\begin{figure}[t]
    \centering
    \captionsetup[subfigure]{labelformat=empty} 
	\begin{subfigure}{0.48\textwidth}
    		\centering
    		\includegraphics[width=\textwidth]{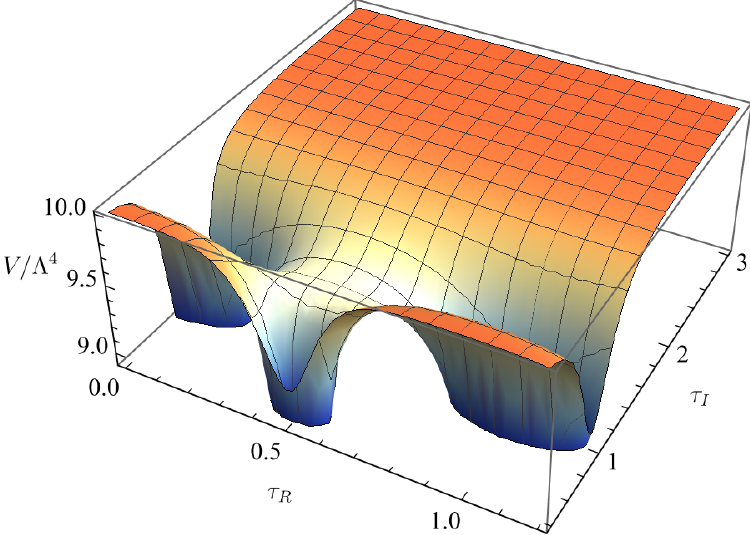}
    		\caption{$\alpha = 0.1$}
	\end{subfigure}
	\quad
	\begin{subfigure}{0.48\textwidth}
		\centering
		\includegraphics[width=\textwidth]{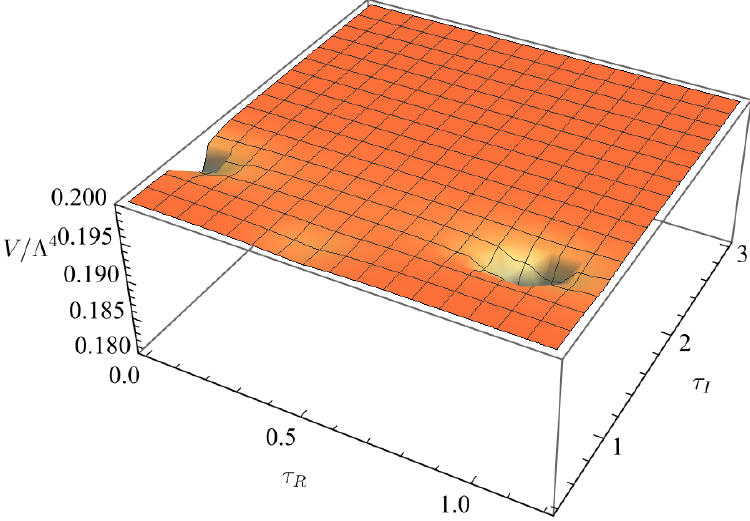}
		\caption{$\alpha = 5$}
	\end{subfigure}
    \caption{The 3D plot of the scalar potential \eqref{eq:model1-potential} with the correction to the K\"{a}hler potential~\eqref{eq:model1-deltaK}.}
    \label{fig:model1-potential}
\end{figure}

\medskip
As discussed in the details in the next section\footnote{
See also App.~\ref{app:Y^{3,6}-and-Y^{3,4}}, in which
a model with two kinds of modular forms in the superpotential
is discussed in the spirit of multi-natural inflation \cite{Czerny:2014wza,Czerny:2014xja}. Then, the slow-roll inflation seems difficult to be realized due to the moduli destabilization during the inflation.}, this simple potential
\eqref{eq:model0-potential}
does not have a flat direction enough to realize the slow-roll inflation.
In order to obtain the slow-roll inflation, 
let us introduce the following additional term to the K\"{a}hler potential in addition to \eqref{eq:model0-Kahlerpotential}
\begin{align}
	\delta \mc{K} =  \alpha \frac{|Y^{3,6}_{\mb{1}} X|^2}{(- \iu \tau + \iu \bar{\tau})^{-k}}
  =  \alpha \frac{|Y^{3,6}_{\mb{1}} X|^2}{(- \iu \tau + \iu \bar{\tau})^{h}},
	\label{eq:model1-deltaK}
\end{align}
where $k$ is the modular weight associated with this operator and $\alpha$ is a positive and dimensionless constant characterizing this additional term.
The existence of this kind of term is discussed in Refs.~\cite{Antoniadis:1992pm,Chen:2019ewa,Nilles:2020kgo,Baur:2022hma}.\footnote{
When $\alpha$ is small such a $\delta K$ would be supposed to be generated by radiative corrections. The modular form can be regarded as the Yukawa coupling of $X$ to the heavy modes and hence it is natural for $Y$ to appear in the wave function renormalization of $X$.
}
The modular weight $k$ satisfies $k = k_Y + k_X=-h=-2$, 
because the superpotential
$W \sim Y^{3,6}_{\mb{1}} X$ has the modular weight $-h=-2$.
Then, in the presence of $\delta \mc{K}_{X \bar{X}}$,
the scalar potential \eqref{eq:model0-potential} 
is deformed as 
\begin{align}
	V = \Lambda^4 \frac{(2 \tau_I)^6 |Y^{3,6}_{\bm{1}}(\tau)|^2}{1 + \alpha (2 \tau_I)^6 |Y^{3,6}_{\bm{1}}(\tau)|^2}.
    \label{eq:model1-potential}
\end{align}
If we write 
\begin{align}
	\tilde{V} = (2\tau_I)^6 |Y^{3,6}_{\bm{1}}(\tau)|^2 \geq 0,
\end{align}
this potential is written as 
\begin{align}
	V = \Lambda^4 \frac{\tilde{V}}{1 + \alpha \tilde{V}}.
\end{align}
This shows that the vacuum is determined by the modular form.
Note that the vacuum of $V$ is basically the same as that of $\tilde{V}$, which shares similarity to Refs.~\cite{Kobayashi:2016mzg,Kobayashi:2019uyt}.
The scalar potential is in general suppressed by
$1/(1+\alpha \tilde{V})$ for $\alpha >0$ and hence can get flatter enough to realize the slow-roll inflation. 
In the $\tilde{V}$ dominate region, this potential reduces to the homogeneous constant profile, $V \sim \Lambda^4 / \alpha$. 
The shapes of the potentials are shown in Fig.~\ref{fig:model1-potential}, where $\alpha = 0.1$ in the left panel and $\alpha = 5$ in the right panel.
If $\alpha$ becomes large, the potential behaves as the constant one $\sim \Lambda^4 / \alpha$ except the minimal points of $\tilde{V}$.
The potential has a pinhole like minimum as $\alpha$ becomes larger. This feature will be similar to the $\alpha$-attractor models \cite{Kallosh:2013hoa,Kallosh:2013yoa,Galante:2014ifa,Kallosh:2014rga,Kallosh:2015lwa,Linde:2015uga,Carrasco:2015pla}. 
Vacuum in the potential is the CP conserving and hence the modulus tends to be stabilized in the CP-symmetric vacuum for the weight 6 modular form case.\footnote{In the top-down approach to stabilize the moduli fields, the CP symmetry is also preserved at the vacua \cite{Kobayashi:2020uaj,Ishiguro:2020nuf}.}

\paragraph{Comments on axion weak gravity conjecture.}

Note that there exists $\tau_R$ which is shifted by the $T$ transformation and hence known to behave like an axion. 
Let us comment on the constraint from axion weak gravity conjecture~\cite{Arkani-Hamed:2006emk}, which claims that the axion decay constant $f_a$ should satisfy 
\begin{align}
    \bm{S}_{\mr{inst}} \lesssim \mc{O}(1) \frac{M_P}{f_a},
    \label{eq:aWGC}
\end{align}
where $\bm{S}_{\mr{inst}}$ denotes an instanton action determining the normalization of the axion potential by $V \sim \ee^{-\bm{S}_{\mr{inst}}}$.
The scalar potential associated with the modular invariance is given by $V \sim q^n + \bar{q}^n$, where$q = \exp ( 2 \pi \iu \tau)$ and $n \in \mbb{R}^+$, then the instanton action reads
\begin{align}
    \bm{S}_{\mr{inst}} \sim - n \log (|q|) = 2 \pi n \braket{\tau_I}.
\end{align}
The kinetic term of the axion is 
\begin{align}
    \bm{L}_{\mr{kin}} = - \frac{h M_P^2}{4 \braket{\tau_I}^2} (\del_\mu \tau_R)^2 \eqqcolon - \frac{1}{2}(\del_\mu a)^2, 
\end{align}
from Eqs.~\eqref{eq:Kahler-potential-tau} and \eqref{eq:Kahler-metric-tau}, where the canonically normalized axion is defined by $a = \sqrt{\frac{h}{2}} \frac{M_P}{\braket{\tau_I}} \tau_R$.
From the periodicity of $a \simeq a + 2 \pi f_a$, the axion decay constant reads 
\begin{align}
    f_a \sim \frac{M_P}{2 \pi n \braket{\tau_I}} \sqrt{\frac{h}{2}}.
\end{align}
Thus, the inequality of axion weak gravity conjecture \eqref{eq:aWGC} becomes 
\begin{align}
    \bm{S}_{\mr{inst}} \sim  2\pi n \braket{\tau_I} \lesssim \frac{M_P}{f_a} \sim 2 \pi n \braket{\tau_I} \sqrt{\frac{2}{h}},
\end{align}
and we find 
\begin{align}
    \sqrt{\frac{h}{2}} \lesssim \mc{O}(1),
\end{align}
which is automatically satisfied for $h = 1,2,3$.

\subsection{Toy model}

In this subsection, let us make progress of the intuitive understanding of the inflation via the potential \eqref{eq:model1-potential} by using toy models inspired from the $q$ expansion.
From the $q$ expansion result \eqref{eq:q-expansion-Y^{3,6}_1}, the modular form has roughly the structure of $Y \sim 1 - B \ee^{ 2\pi \iu \tau}$, where $B$ is a $\tau$ independent constant.
Using this and the SUGRA formula, let us show how the flat direction for the slow-roll inflation is produced by the additional contribution by $\alpha$ in the $\tau_R$ direction and $\tau_I$ direction, respectively.
The scalar potential $V \sim \tilde{V}/(1+\alpha \tilde{V})$
is suppressed by the denominator for $\alpha >0$ and hence can become flatter for realizing the successful slow-roll inflation.

\paragraph{$\tau_R$ direction inflation.}

\begin{figure}[t]
	\centering
	\includegraphics[width=0.48\textwidth]{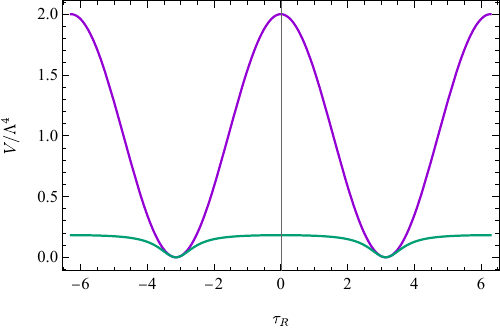}
	\caption{The $\tau_R$ potential \eqref{eq:potential-tauR-toy}.
	We choose $\alpha = 5$ and $A =1$ for the flatter green line, and $\alpha = 0$ and $A =1$ for the steeper purple line.}
	\label{fig:toy-model-potential-tauR}
\end{figure}

In this case, the scalar potential is given by 
\begin{align}
	V = \Lambda^4 \frac{\tilde{V}}{1 + \alpha \tilde{V}},
	\qquad 
	\tilde{V} = \cos ( 2\pi \tau_R) + A,
	\label{eq:potential-tauR-toy}
\end{align}
where we use the SUGRA formula to derive 
$\tilde{V} \sim |Y|^2 \sim \cos (2\pi \tau_R) + A$ 
and
$\alpha$ is a deformation parameter and assumed to be positive as discussed in the previous subsection.
This potential is shown in Fig.~\ref{fig:toy-model-potential-tauR}.
The steeper purple line is $\tilde{V}$ and the flatter green line is the deformed potential $V$.
The potential is suppressed by the additional contribution from $\alpha$, and it is found the flat direction arises between the minima, where the slow-roll inflation can occur.
The existence of the minima does not change by this deformation.
Note that for a large $\alpha$ the potential can become sufficiently flat for the slow-roll inflation even without the decay constant larger than the Planck scale.

\paragraph{$\tau_I$ direction inflation.}

\begin{figure}[t]
	\centering
	\includegraphics[width=0.48\textwidth]{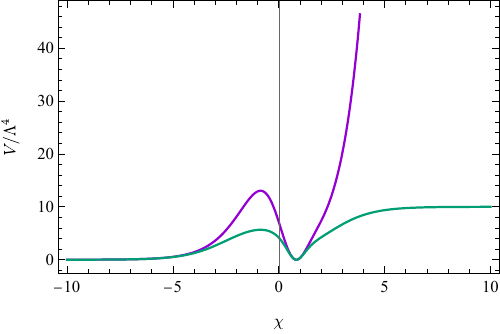}
	\quad
	\includegraphics[width=0.48\textwidth]{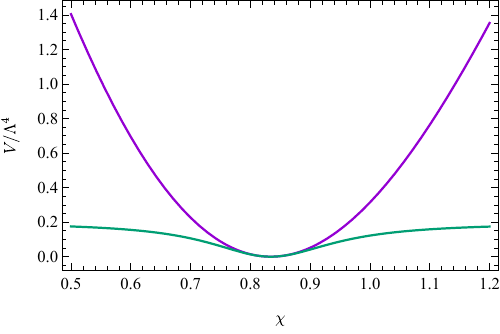}
	\caption{The $\tau_I \sim \ee^{\chi}$ potential \eqref{eq:potential-tauI-toy}.
	We choose $\alpha = 5$ and $A =10$ for the flatter green line, and $\alpha = 0$ and $A =10$ for the steeper purple line. The right panel shows the enlarged view around the vacuum.}
	\label{fig:toy-model-potential-tauI}
\end{figure}

In this case, the scalar potential is given by 
\begin{align}
	V = \Lambda^4 \frac{\tilde{V}}{1 + \alpha \tilde{V}},
	\qquad 
	\tilde{V} = \ee^{\chi} \bigl[ 1 - A \exp ( -\ee^{\chi}) \bigr]^2.
	\label{eq:potential-tauI-toy}
\end{align}
From the SUGRA formula, we calculate $\tilde{V}$ as 
\begin{align}
	\tilde{V} &\sim \tau_I^n |Y|^2 \sim \tau_I^n (1 - A \ee^{- \tau_I} )^2
	\nn\\
	&\sim \ee^{\chi} \bigl[ 1 - A \exp ( -\ee^{\chi}) \bigr]^2,
\end{align}
where $\tau_I \sim \ee^\chi$ and $A \gg 1$ is assumed
and the size of $n$ is irrelevant to this discussion so long as $n={\cal O}(1)$.
This field redefinition is motivated by the non-canonical kinetic term of $\tau_I$, $\bm{L}_{\mr{kin.}} \sim - (\del_\mu \tau_I / \tau_I)^2$.
The potential is shown in Fig.~\ref{fig:toy-model-potential-tauI}.
The steeper purple (flatter green) line corresponds to $\tilde{V}$ ($V$).
As shown in these panels, the scalar potential is pushed down by the additional contribution from $\alpha$ and we find the flat direction around the $\chi$ vacuum.

\section{Modular flavor inflation}
\label{sec:inflation}

We discuss the slow-roll inflationary scenario depending on $\alpha$ in our model.
The formulae of multi-field inflation~\cite{Sasaki:1995aw,Chiba:2008rp,Salvio:2021lka} are used in order to evaluate the slow-roll parameters, power spectrum, spectral index, and tensor-to-scalar ratio.
The slow-roll parameters are given by 
\begin{align}
	\epsilon_V = \frac{V_{,a} V^{,a}}{2 V^2},
	\qquad
	\eta_V{}^a_b = \frac{V^{;a}_{;b}}{V},
\end{align}
where $_{,a}$ and $_{;a}$ denote the derivative and covariant derivative with respect to $\phi^a$, respectively.
The Levi-Civita connection for this covariant derivative is calculated from the metric $K_{ab}$, and
$\eta_V{}^a_b$ becomes a matrix in a multi-field inflation.
The power spectrum $\mc{P}_{\mc{R}}$, spectral index $n_s$, and tensor-to-scalar ratio $r$ are expressed as 
\begin{align}
	\mc{P}_{\mc{R}} &= \biggl( \frac{H}{2 \pi} \biggr)^2 N_{,a} N^{,a},
	\\
	n_s &= 1 - 2 \epsilon_V - \frac{2}{N_{,a} N^{,a}} + \frac{2 \eta_{Vab} N^{,a} N^{,b}}{N_{,c}N^{,c}},
	\\
	r &= \frac{8}{N_{,a} N^{,a}},
\end{align}
where $H$ denotes the Hubble parameter during the inflation.
Here, $N_{,a}$ is given by\footnote{
As discussed in App.~\ref{app:multi-field-inflation}, in this work we do not include contribution of the isocurvature fluctuation, which is orthogonal to the adiabatic fluctuation on the inflationary trajectories.
}
\begin{align}
	N_{,a} = \frac{V V_{,a}}{V_{,b} V^{,b}}.
\end{align}
The e-folding is defined by $N = \log (a_f/a)$,
where $a_f$ is the scale factor at the end of the inflation and $a$ is the one at e-folding $N$ before the end of inflation.
For the more details, see App.~\ref{app:multi-field-inflation}.

\subsection{Simple model with $\alpha =0$}

\begin{figure}[t]
	\centering
	\includegraphics[width=0.48\textwidth]{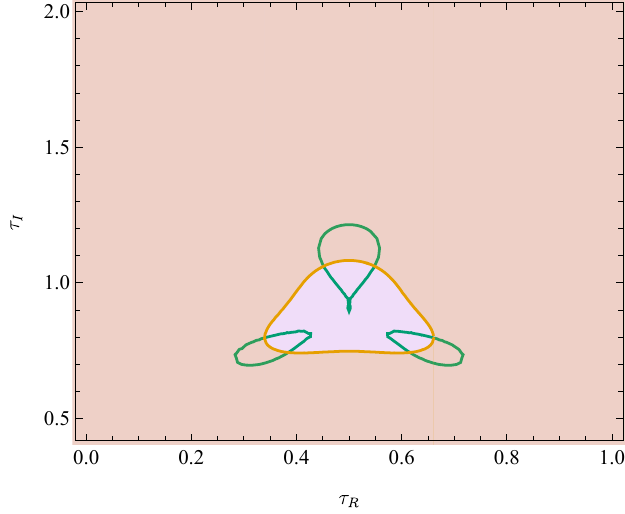}
	\caption{The parameter space for the slow-roll inflation with the potential \eqref{eq:model0-potential}. The slow-roll inflation does not take place in this model.
	See the texts for the details of each region.}
	\label{fig:model0-taunssrp}
\end{figure}

First, let us start with the discussion about the simple model with the scalar potential \eqref{eq:model0-potential}, which is an undeformed one.
Fig.~\ref{fig:model0-taunssrp} shows the parameter spaces
for the slow-roll inflation in $(\tau_R, \tau_I)$-plane, where it turns out that the slow-roll inflation does not take place in this simple model. Here, the colored regions indicate the breakdown of the slow-roll condition: the purple region shows $|\eta_V{}^a_b | \geq 1$, whereas the orange one shows $\epsilon_V \geq 1$.
The boundary $\epsilon_V =1$ is set to the field values at the end of the inflation $(N=0)$ throughout this paper.
On the green line, we obtain $n_s = 0.965$ which is the
central value of the current observation~\cite{Planck:2018vyg}.

\subsection{Deformed model with $\alpha \neq 0$}

In this subsection, we study the parameter space for the successful slow-roll inflation based on the scalar potential~\eqref{eq:model1-potential} with the kinetic term 
\eqref{eq:kinetic-term-modulus}.
The correction to the K\"{a}hler potential~\eqref{eq:model1-deltaK} flattens the scalar potential, and hence
there indeed exists the parameter space in which the slow-roll inflation can take place.
As shown below, for $\alpha \lesssim 1$, 
$\tau_I$ behaves as the inflaton of the slow-roll inflation,
which lasts for a sufficiently long time. Then $\tau_R$ can play a role of the waterfall field of the so-called hybrid inflation at the end of the inflation and settles down into the CP-conserving vacuum at last, when $\tau_R$ develops a non-zero value during the inflation.
For $\alpha \gtrsim 1$, a combination of $\tau_R$ and $\tau_I$
plays a role of the inflaton, since the scalar potential has the homogeneous constant profile $\sim \Lambda^4/\alpha$ when the slow-roll inflation occurs in apart from the vacuum.
It turns out that $\tau_I$ can become the inflaton in terms of the pole inflation \cite{Broy:2015qna,Terada:2016nqg} for any $\alpha \gtrsim 10^{-2}$.

\begin{figure}[t]
    \centering 
    \captionsetup[subfigure]{labelformat=empty} 
    \begin{subfigure}{0.48\textwidth}
        \centering
        \includegraphics[width=\textwidth]{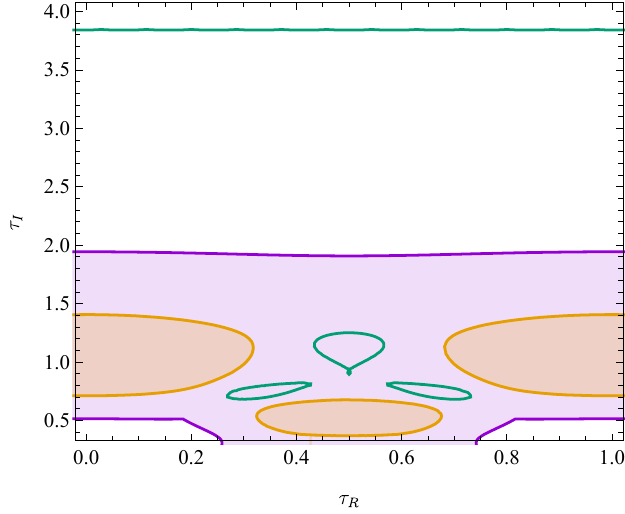}
        \caption{$\alpha = 0.01$}
    \end{subfigure}
    \quad
    \begin{subfigure}{0.48\textwidth}
        \centering
        \includegraphics[width=\textwidth]{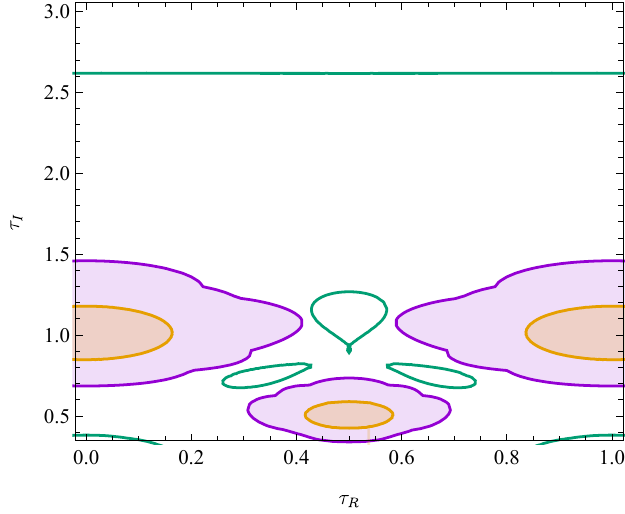}
        \caption{$\alpha = 0.1$}
    \end{subfigure}
    \vspace{2ex}
    \\
    \begin{subfigure}{0.48\textwidth}
        \centering
        \includegraphics[width=\textwidth]{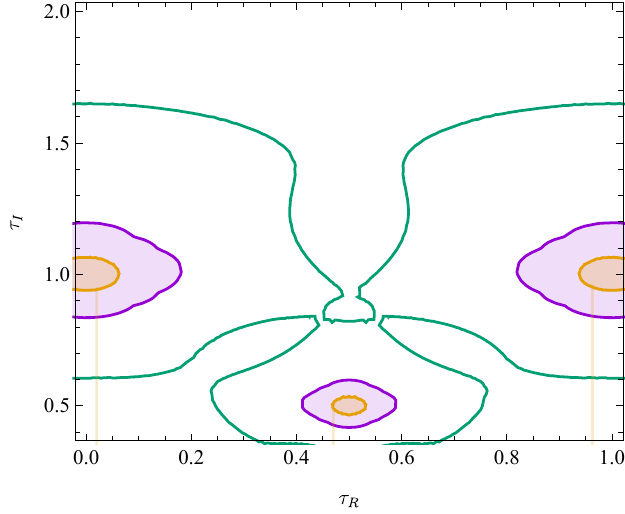}
        \caption{$\alpha = 2$}
    \end{subfigure}
    \quad
    \begin{subfigure}{0.48\textwidth}
        \centering
        \includegraphics[width=\textwidth]{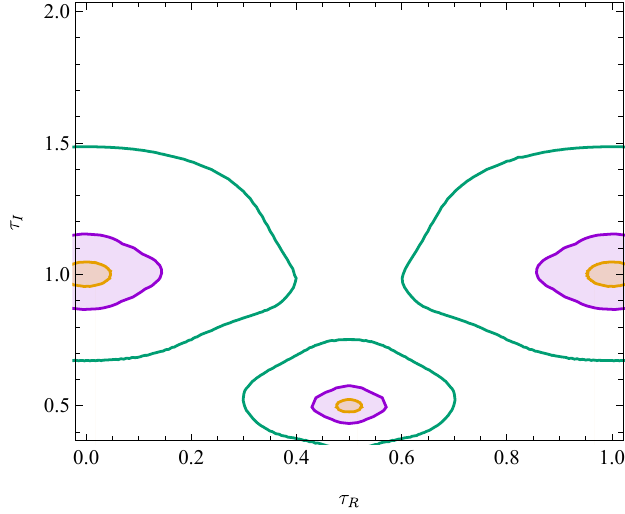}
        \caption{$\alpha = 5$}
    \end{subfigure}
    \caption{The parameter space for the slow-roll inflation with \eqref{eq:model1-potential} including the correction to the K\"{a}hler potential \eqref{eq:model1-deltaK}.
    We choose $\alpha = 0.01,\ 0.1,\ 2,$ and 5.
    }
    \label{fig:model1-tausrpns-alp}
\end{figure}

\begin{figure}[t]
    \centering
    \begin{subfigure}{0.48\textwidth}
        \centering
        \includegraphics[width=\textwidth]{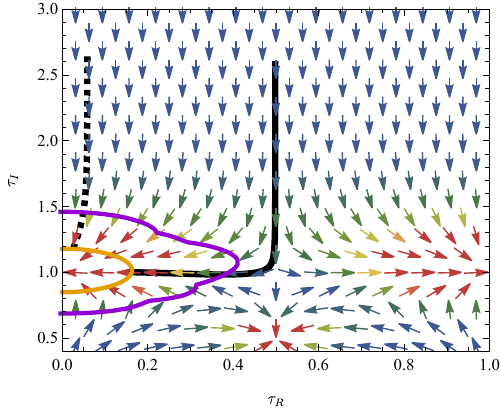}
        \caption{$\alpha = 0.1$}
    \end{subfigure}
    \quad 
    \begin{subfigure}{0.48\textwidth}
        \centering
        \includegraphics[width=\textwidth]{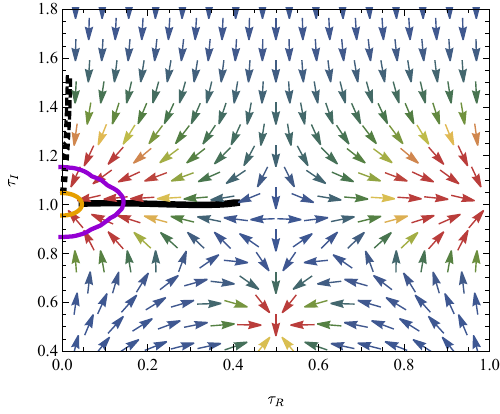}
        \caption{$\alpha = 5$}
    \end{subfigure}
    \caption{The vector plots of the potential gradients $- \frac{K^{ab} V_{,b}}{V}$.
    The black solid and dashed lines are inflationary trajectories starting at e-folding $N=60$ before the end of the inflation.
    Orange and purple contours show $\epsilon_V =1$ and $|\eta_V{}^a_b| =1$ respectively.
    As for the black bold line in the left panel, the roll of $\tau_R$ at the end of the inflation violates slow-roll condition and $\tau_R$ plays the roll of the waterfall field in the hybrid inflation.
    }
	\label{fig:model1-vectraj-alp}
\end{figure}

\medskip
Fig.~\ref{fig:model1-tausrpns-alp} shows the parameter space 
for the slow-roll inflation
in $(\tau_R, \tau_I)$-plane in models with $\alpha \neq 0$.
The meanings of each colored region and green lines are the same as those in Fig.~\ref{fig:model0-taunssrp} and hence the blank region in the Fig.~\ref{fig:model1-tausrpns-alp} implies that 
there exist successful inflationary trajectories for the slow-roll inflation
which lasts for a long time.\footnote{
The blank region shows that all slow-roll parameters are smaller than unity throughout this paper.
Then mode orthogonal to the inflaton is also light,
however, isocurvature mode is not discussed in this paper.
}
It is noted that the contribution of $\alpha$ changes the parameter space for the successful slow-roll inflation, i.e., a candidate of the inflaton.
See also Fig.~\ref{fig:model1-vectraj-alp}, which shows the vector plots of the potential gradient $- \frac{K^{ab} V_{,b}}{V}$ and examples of the inflationary trajectories starting at $N = 60$ denoted by bold lines and dashed ones. In the slow-roll regime, where $\epsilon_V <1,~|\eta_V{}^a_b|<1$, 
the equations of motion (EOMs) of $\tau$ \footnote{
The full EOMs of $\tau$ are discussed in App.~\ref{app:field-equations}.
} are approximately given by 
\begin{align}
    \tau'_R - \frac{2 \tau_I^2  V_{,\tau_R}}{h V} = 0,
    \qquad 
    \tau'_I - \frac{2 \tau_I^2 V_{,\tau_I}}{h V} =0,
    \label{eq:tau-EOMs-inflation}
\end{align}
and the inflaton moves along this potential gradient associated with Fig. \ref{fig:model1-vectraj-alp}.
Prime denotes the derivative with respect to the e-folding $N$.
$\tau_I^2$ in the second term in the both EOMs comes from the inverse scalar field metric $K^{ab}$ \eqref{eq:Kahler-metric-tau}.
For $\alpha \lesssim 1$ (upper panels of Fig.~\ref{fig:model1-tausrpns-alp}), there exists a green line at a larger $\tau_I = {\rm const.}$ and wide colored regions around a smaller $\tau_I$ where the slow-roll condition is violated. In the left panel of Fig.~\ref{fig:model1-vectraj-alp} we find inflationary trajectories, which are actually present within the blank region in the Fig.~\ref{fig:model1-tausrpns-alp}. As arrows shown in the left panel of Fig.~\ref{fig:model1-vectraj-alp}, the inflation turns out to be mainly driven by $\tau_I$.
The inflaton $\tau_I$ starts to roll from a larger value to smaller one, and the field value realizing $n_s = 0.965$ at the horizon exit is given by the green line of $\tau_I = {\rm const.}$ and the slow-roll inflation ends at the orange contour as shown 
in Figs. \ref{fig:model1-tausrpns-alp} and \ref{fig:model1-vectraj-alp}.
In the left panel of Fig. \ref{fig:model1-vectraj-alp}, the solid line 
starting from $\tau = 1/2 + \iu$ to $\tau =\iu$ along $\tau_R$ direction shows the last stage of the inflation, where the slow-roll condition is violated.
Then, the inflationary energy along $\tau_I$ direction converts to that of $\tau_R$, hence the inflation ends and moduli settles down to the CP-conserving vacuum.\footnote{
The constraint on isocurvature fluctuation could give conditions to our model because the modulus field perpendicular to the inflaton on the inflationary trajectories can be also lighter than the Hubble scale during the inflation in our model.
However, in this work, we do not study this constraint further.
}
This is regarded as a kind of the hybrid inflation and
$\tau_R$ is then the waterfall field for it.
As $\alpha$ becomes larger (in the bottom panels of Fig.~\ref{fig:model1-tausrpns-alp}), a green contour at a larger $\tau_I = \mr{const.}$ merges with those around $\tau_I \sim 1$, and there appear green contours around the stationary points in the scalar potential. 
The scalar potential \eqref{eq:model1-potential} has the pinhole-like shaped vacua due to the deformation by $\delta \mc{K}$ as discussed in Sec.~\ref{sec:model}, and the slow-roll inflation ends at the orange contour around the vacua.
(Note that the vacuum at $\tau_R = 0$ is identified with that at $\tau_R = 1$.)
Inflationary trajectories are allowed to exist within the wider blank region in the bottom panels of Fig.~\ref{fig:model1-tausrpns-alp} than that for a smaller $\alpha$, 
and green contours show the variety of the field values at the horizon exit. Thus, in general, a combination of $\tau_R$ and $\tau_I$ is thought to be the inflaton.
For instance, either $\tau_R$ or $\tau_I$ can drive the single-field inflation as seen in the right panel of Fig.~\ref{fig:model1-vectraj-alp}.
Note that for a large $\alpha$ the potential along $\tau_R$ direction $V \sim \cos(\tau_R)/[1+\alpha (\cos(\tau_R)+A)]$ becomes sufficiently flat for the slow-roll inflation even without a larger decay constant than the Planck scale as discussed in the toy model.
We note also that $\tau_I$ drives the inflation in terms of the pole inflation \cite{Broy:2015qna,Terada:2016nqg} around $\tau_R \sim 0$ for $\alpha \gtrsim 10^{-2}$.

\medskip
We find that the scalar potential has another CP-conserving vacuum at $\tau = \frac{1 + \iu}{2}$ as shown in the left panel of Fig.~\ref{fig:model1-potential}. Figs. \ref{fig:model1-tausrpns-alp} and \ref{fig:model1-vectraj-alp} also indicate the presence of the vacuum.
The moduli can be stabilized at this vacuum at the end of the inflation when the inflaton starts to roll in the region where $\tau_I \sim 1/2$.
However, $\tau = \frac{1 + \iu}{2}$ is identical to $\tau = \iu$ under the $S$ and $T$ transformations. 

\begin{figure}[t]
    \centering
    \begin{subfigure}{0.48\textwidth}
        \centering
        \includegraphics[width=\textwidth]{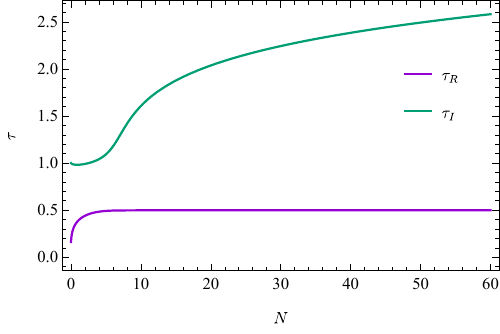}
        \\
        \includegraphics[width=\textwidth]{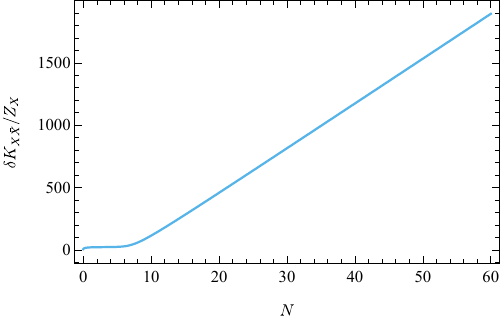}
        \caption{$\alpha = 0.1$}
    \end{subfigure}
    \quad 
    \begin{subfigure}{0.48\textwidth}
        \centering
        \includegraphics[width=\textwidth]{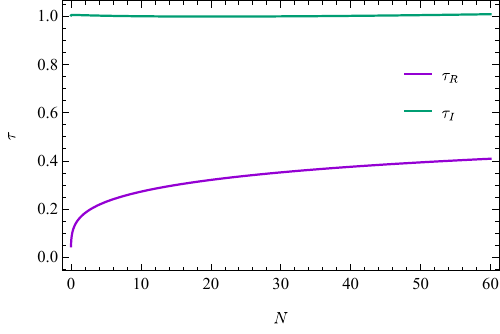}
        \\
        \includegraphics[width=\textwidth]{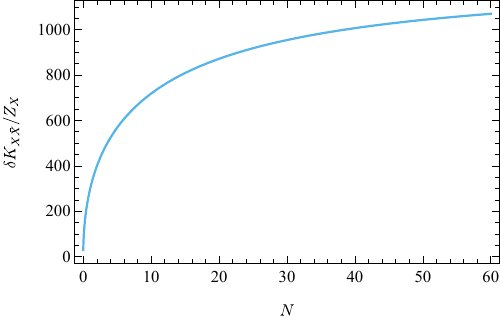}
        \caption{$\alpha = 5$}
    \end{subfigure}
    \caption{The time evolution of the moduli fields along the black solid lines in Fig.~\ref{fig:model1-vectraj-alp}.
    $\alpha$ is chosen as $0.1$ (left) and $5$ (right).
	$N$ denotes the e-folding before the end of the inflation.
    The purple and green lines are the profiles for $\tau_R$ and $\tau_I$, respectively.
    In the lower panels, we show the time evolution of $\delta \mc{K}_{X\bar{X}}/Z_X = \alpha |Y^{3,6}_{\bm{1}}|^2 (2 \tau_I)^{k - k_X} = 
    \alpha |Y^{3,6}_{\bm{1}}|^2 (2 \tau_I)^{6} $ with $k=-h=-2$ and $k_X=-8$.
	}
	\label{fig:model1-Ntau-p1}
\end{figure}
\begin{figure}[ht]
	\centering
    \renewcommand\thesubfigure{\alph{subfigure}$^{\prime}$}
    \begin{subfigure}{0.48\textwidth}
        \centering
        \includegraphics[width=\textwidth]{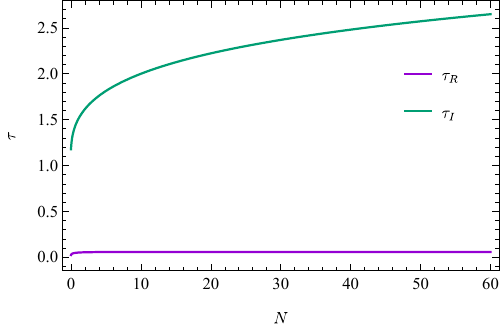}
        \\
        \includegraphics[width=\textwidth]{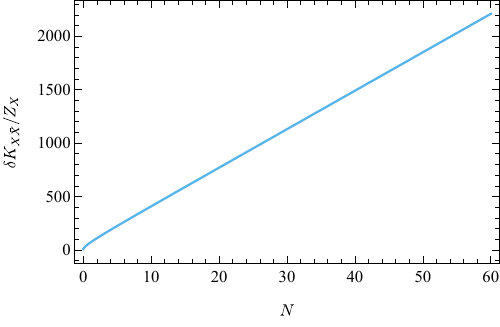}
        \caption{$\alpha = 0.1$}
    \end{subfigure}
    \quad 
    \begin{subfigure}{0.48\textwidth}
        \centering
        \includegraphics[width=\textwidth]{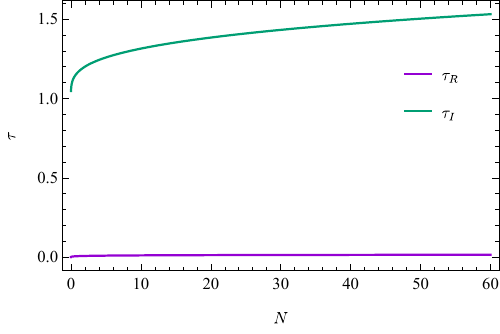}
        \\
        \includegraphics[width=\textwidth]{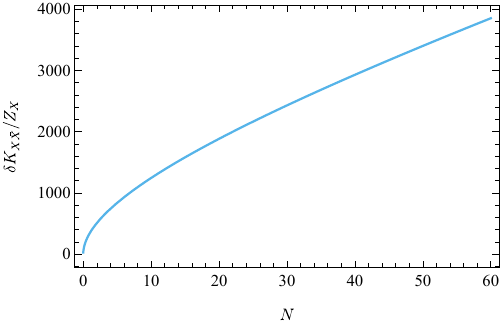}
        \caption{$\alpha = 5$}
    \end{subfigure}
    \caption{The time evolution of the moduli fields along the black dashed lines in Fig.~\ref{fig:model1-vectraj-alp}.
    See the caption of Fig.~\ref{fig:model1-Ntau-p1} for the details.
    }
    \label{fig:model1-Ntau-p2}
\end{figure}

\medskip
In the following part, we will mainly focus on the inflationary trajectories from the field values at the e-folding $N=60$ to those around the vacuum $(\tau_R, \tau_I) = (0,1)$ at $N=0$.
Fig.~\ref{fig:model1-Ntau-p1} shows the time evolution of $\tau$ (upper panels) and 
the magnitude of $\delta \mc{K}_{X \bar{X}}$ with $k=-h=-2$ (lower panels) in terms of e-folding $N$ along the black solid lines in Fig.~\ref{fig:model1-vectraj-alp}.
Fig.~\ref{fig:model1-Ntau-p2} is similar but shows those on the dashed lines in Fig.~\ref{fig:model1-vectraj-alp}.
In the upper panels of both Figs.~\ref{fig:model1-Ntau-p1} and \ref{fig:model1-Ntau-p2}, the purple and green lines give the profile of $\tau_R(N)$ and $\tau_I(N)$ respectively.
In the upper left panel of Fig.~\ref{fig:model1-Ntau-p1}, 
inflaton $\tau_I$ rolls from a large value to smaller one, while $\tau_R$ stays steady at $\tau_R = 1/2$ during the inflation for $N\gtrsim 1$ and starts to roll at $N \sim 1$ as the waterfall field in the hybrid inflation at the late stage of the inflation.
Then $\tau$ settles down into the vacuum $\tau = \iu$.
Note that for $\alpha \lesssim 1$ the adiabatic perturbation of $\tau_I$ 
at the horizon exit (as the coordinate around $\tau \sim 1/2 + 2.6 \iu$ for $\alpha = 0.1$) can realize the spectral index consistent with the current observation.
In the upper right panel of Fig.~\ref{fig:model1-Ntau-p1}, 
on the other hand, the inflaton $\tau_R$ rolls from a large value to smaller ones, while $\tau_I$ gets remain at $\tau_I = 1$.
Then, the adiabatic perturbation of $\tau_R$ at the horizon exit (as the coordinate around $\tau \sim 0.4 + \iu$ for $\alpha = 5$) can realize the consistent $n_s$ with the current observation.
In the lower panels of Fig.~\ref{fig:model1-Ntau-p1},
$\delta \mc{K}_{X \bar{X}}$ is shown to be smaller than the leading field metric of $X$, $Z_X \coloneqq 1/(2\tau_I)^{-k_X}$ with $k_X=-8$, at the vacuum $\tau = \iu$, but can be larger than $Z_X$ during the inflation even though $\alpha < 1$ can naively be regarded as a perturbative correction.
Thus, $\delta \mc{K}$ might not be regarded as the mere perturbative correction to the field metric during the inflation and be originated non-perturbatively from a strong coupling.
Therefore, issue of controlling models could arise against modular forms which might exist in our models.
However, throughout this paper, it is assumed that coefficients of such modular forms in the action are suppressed and dynamics of $X$ is stabilized, and hence we will not discuss this issue further.

\medskip
As shown in the upper panels of Fig.~\ref{fig:model1-Ntau-p2}, the inflation is mainly driven by $\tau_I$ while $\tau_R$ almost gets remain at $\tau_R \sim 0$.
The lower panels show the $N$ dependence of $\delta \mc{K}_{X\bar{X}}$ by using the these $\tau$ profiles.
A similar issue concerned with the modular forms in our models could arise
as in the previous case. 
See also Fig.~\ref{fig:model1-Ntau-end} in  App.~\ref{app:field-equations}, which shows the time evolution of the moduli after the slow-roll inflation on the black solid lines in Fig.~\ref{fig:model1-vectraj-alp}.
Moduli settle down into the vacuum immediately after the end of the slow-roll inflation, oscillating around the vacuum.

\begin{figure}
    \centering 
    \begin{subfigure}{0.48\textwidth}
        \centering 
        \includegraphics[width=\textwidth]{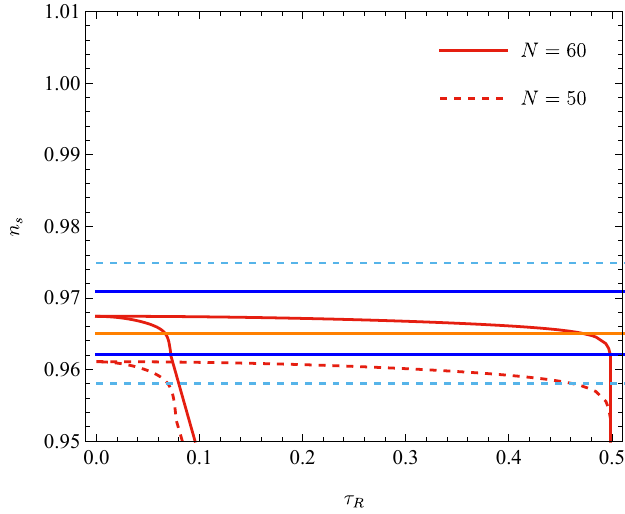}
        \\
        \includegraphics[width=\textwidth]{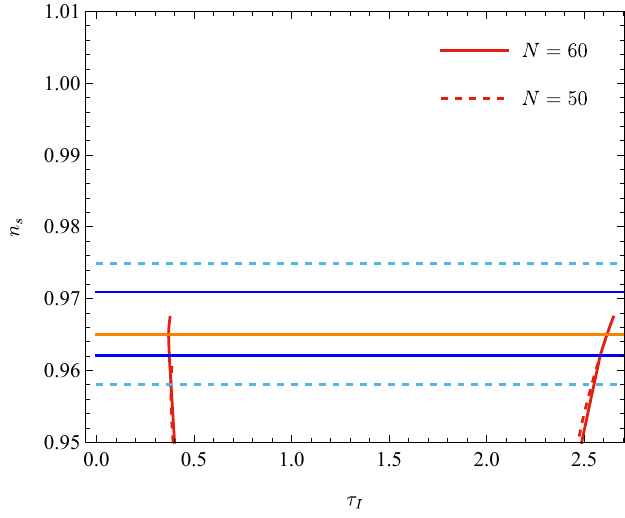}
        \caption{$\alpha = 0.1$}
    \end{subfigure}
    \quad
    \begin{subfigure}{0.48\textwidth}
        \centering
        \includegraphics[width=\textwidth]{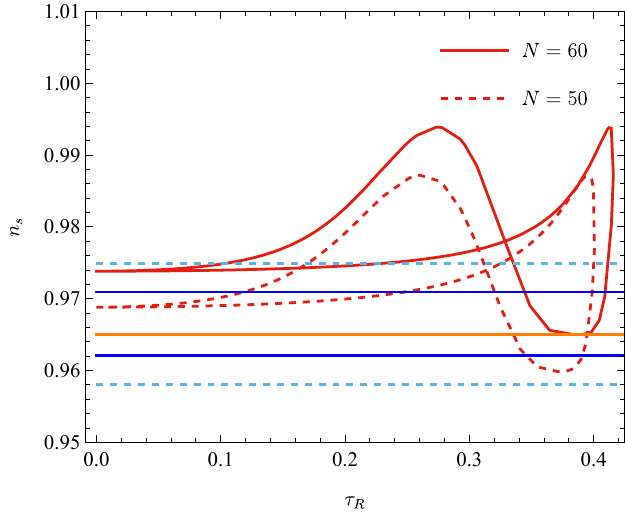}
        \\
        \includegraphics[width=\textwidth]{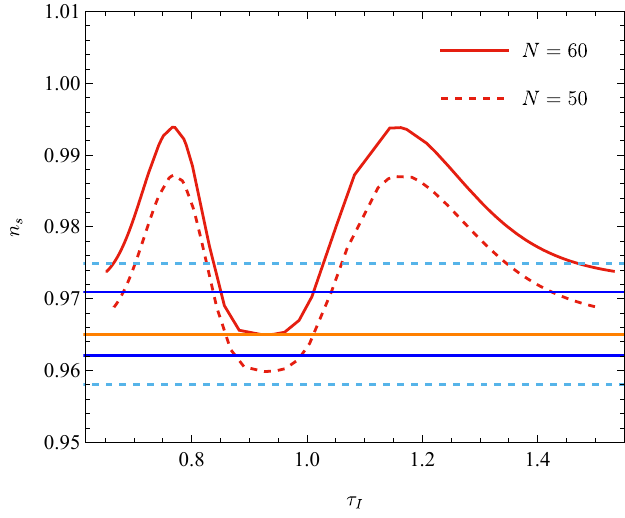}
        \caption{$\alpha = 5$}
    \end{subfigure}
    \caption{
        Predictions for $n_s$ in our models for $\alpha=0.1$ and 5.
    }
    \label{fig:model1-tauRIns-alp}
\end{figure}
\begin{figure}
    \centering 
    \begin{subfigure}{0.48\textwidth}
        \centering 
        \includegraphics[width=\textwidth]{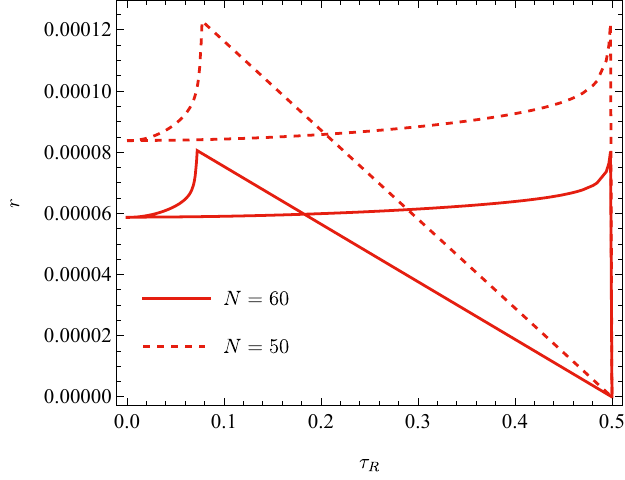}
        \\
        \includegraphics[width=\textwidth]{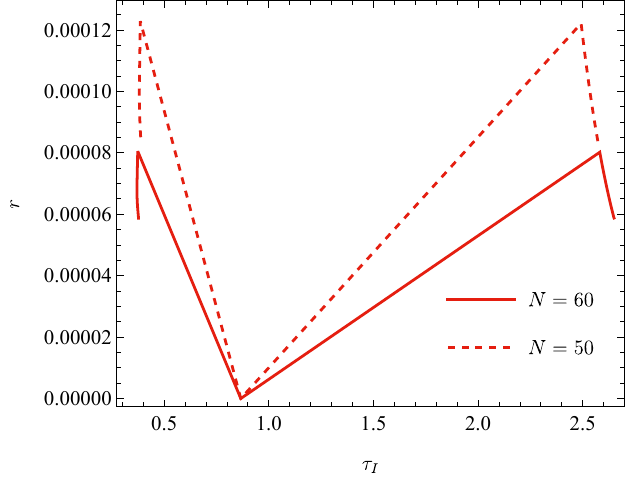}
        \caption{$\alpha = 0.1$}
    \end{subfigure}
    \quad
    \begin{subfigure}{0.48\textwidth}
        \centering
        \includegraphics[width=\textwidth]{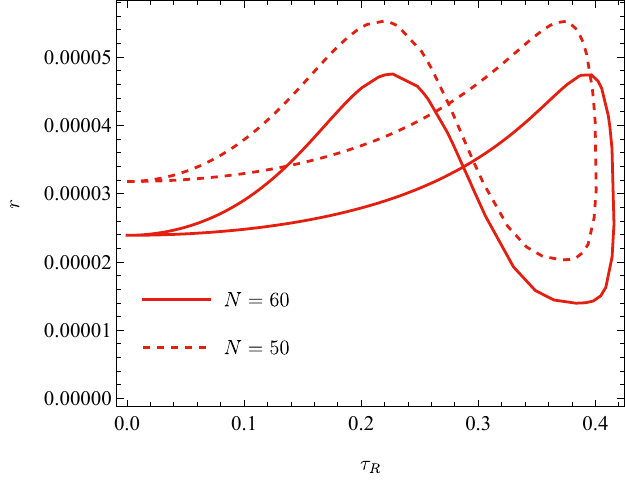}
        \\
        \includegraphics[width=\textwidth]{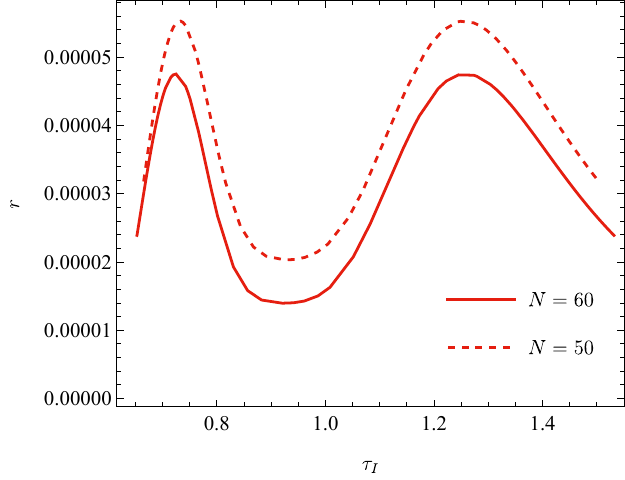}
        \caption{$\alpha = 5$}
    \end{subfigure}
    \caption{
        Predictions for $r$ in our models for $\alpha=0.1$ and 5.
    }
    \label{fig:model1-tauRIr-alp}
\end{figure}

\medskip
Fig.~\ref{fig:model1-tauRIns-alp} shows the moduli-dependence of the
spectral index $n_s$ at $N=60$ (red solid line) and at $N=50$ (red dashed line).
The upper (lower) panels show $\tau_R$- ($\tau_I$)-dependence of $n_s$.
The results for $\alpha = 0.1$ (5) are shown in the left (right) panels.
The orange lines correspond to the current central value, $n_s = 0.965$, 
observed by the PLANCK collaboration~\cite{Planck:2018vyg}.
The blue bold lines (light blue dashed lines) show the $1 \sigma$ ($2 \sigma$) deviation from the central value of the $n_s$. 
It is found that the value of $n_s$ tends to increase as $N$ increases.
As already mentioned above, for $\alpha =0.1$ the inflaton is $\tau_I$ starting to roll from $\tau_I \sim 2.6$ in terms of the pole inflation with a fixed $\tau_R$.
Note also that for $\alpha=5$ a combination of $\tau_R$ and $\tau_I$ drives the inflation. For instance, around $\tau \sim 0.4 + \iu$ ($\tau \sim 1.5\iu$) 
the single-field inflation can be driven by $\tau_R$ ($\tau_I$) with the fixed $\tau_I\sim 1$ ($\tau_R \sim 0$) and these cases will be well-fitted to the current observation.
Similar plots for the tensor-to-scalar ratio $r$ are found in Fig.~\ref{fig:model1-tauRIr-alp}.
In our model, $r$ is tiny and therefore the current constraint, $r < 0.06$, \cite{Planck:2018vyg} is satisfied.

\begin{figure}[t]
	\centering
	\includegraphics[width=0.51\textwidth]{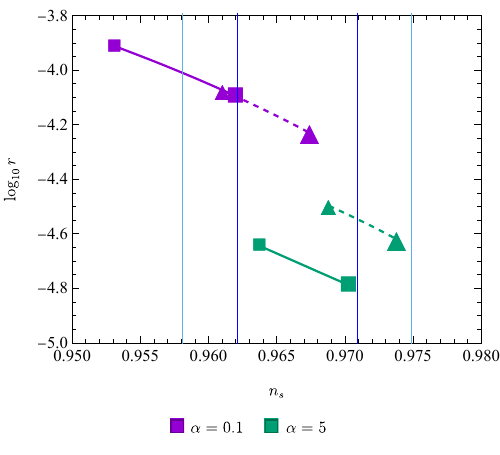}
	\caption{Predictions of our models for the spectral index $n_s$ and the tensor-to-scalar ratio $r$ in the $(n_s, r)$-plane.
	Purple and green lines with marks are results for $\alpha = 0.1$ and 5, respectively.
	The solid (dashed) lines with the squares (triangles) are predictions corresponding to the solid (dashed) trajectories in Fig.~\ref{fig:model1-vectraj-alp}.
	The larger and smaller symbols show the results at $N=60$ and $N=50$ respectively.
	The region surrounded by the blue lines and light blue ones are allowed by the PLANCK observation at $1 \sigma$ and $2 \sigma$ confidence level, respectively.
	}
	\label{fig:model1-nsr-alpcomp}
\end{figure}

\medskip

Fig.~\ref{fig:model1-nsr-alpcomp} shows the predictions of the spectral index $n_s$ and the tensor-to-scalar-ratio $r$ in our models in $(n_s,r)$-plane.
The solid lines with the squares at both ends (the dashed ones with the triangles at both ends) correspond to predictions on the black solid trajectories (the black dashed ones) in Fig.~\ref{fig:model1-vectraj-alp}.
The purple (green) lines are the results for $\alpha = 0.1$ ($5$).
The smaller marks (squares or triangles) represent the result at $N=50$
and the bigger ones show the result at $N =60$.
The larger e-folding $N$ gets, the larger $n_s$ becomes, as shown in Fig.~\ref{fig:model1-tauRIns-alp}.
Further, $r$ gets smaller as $\alpha$ increases.
This behavior is similar to the $\alpha$-attractor models \cite{Kallosh:2013hoa,Kallosh:2013yoa,Galante:2014ifa,Kallosh:2014rga,Kallosh:2015lwa,Linde:2015uga,Carrasco:2015pla}.

\begin{table}[t]
    \centering
    \begin{tabular}{|c||c|c|c|c|c|} 
        \hline
        trajectory & $n_s$ & $r$ & $\mc{P}_{\mc{R}} / (\Lambda/M_P)^4$ & $\Lambda/M_P$ & $|F_X/M_P^2|$ 
        \\
        \hhline{|=#=|=|=|=|=|}
        (a) in Fig.~\ref{fig:model1-Ntau-p1} & $0.962$ & $8.02 \times 10^{-5}$ & $8.41 \times 10^3$ & $7.07 \times 10^{-4}$ & $1.69 \times 10^{-16}$
        \\
        (a$^\prime$) in Fig.~\ref{fig:model1-Ntau-p2} & $0.967$ & $5.89 \times 10^{-5}$ & $1.14 \times 10^4$ & $6.54 \times 10^{-4}$ & $1.45 \times 10^{-16}$
        \\
        (b) in Fig.~\ref{fig:model1-Ntau-p1} & $0.970$ & $1.63\times 10^{-5}$ & $8.28 \times 10^2$ & $1.26 \times 10^{-3}$ & $5.39 \times 10^{-16}$
        \\
        (b$^\prime$) in Fig.~\ref{fig:model1-Ntau-p2} & $0.974$ & $2.39 \times 10^{-5}$ & $5.65 \times 10^2$ & $1.39 \times 10^{-3}$ & $6.54 \times 10^{-16}$
        \\
        \hline
    \end{tabular}
    \caption{Values of the spectral index $n_s$, tensor-to-scalar-ratio $r$, power spectrum  $\mc{P}_{\mc{R}}$, $\Lambda$ in the superpotential, and F-component of $X$ $F_X$ for the trajectories in Figs.~\ref{fig:model1-Ntau-p1} and \ref{fig:model1-Ntau-p2}.
    $n_s$ and $r$ are evaluated at $N=60$.
    }
    \label{tab:values}
\end{table}

\medskip
In the above discussion, we have not considered the normalization $\Lambda$ of the scalar potential in~\eqref{eq:model1-potential} associated with the power spectrum ${\cal P}_{\cal R}$.
This overall scale is fixed by the condition of the power spectrum at the pivot scale~\cite{Planck:2018vyg}
\begin{align}
    \mc{P}_{\mc{R}} = 2.10 \times 10^{-9}.
\end{align}
Taking this into the account against the four inflationary trajectories in Figs.~\ref{fig:model1-Ntau-p1} and \ref{fig:model1-Ntau-p2}, we have fixed the overall scale $\Lambda$ and exhibited it in Tab.~\ref{tab:values}, where $n_s$ and $r$ are also shown.
From the Tab.~\ref{tab:values}, we read $\Lambda \sim 10^{15}$ GeV\footnote{ 
The reduced Planck scale has been shown explicitly in the Tab.~\ref{tab:values}
}. Note that $\Lambda$ are almost independent of models since 
$\mc{P}_{\mc{R}} \sim \Lambda^4/r$ is fixed and $r$ does not drastically change in models.
Let us mention the SUSY breaking scale.
Suppose that stabilizer $X$ breaks the SUSY in the vacuum at $\tau = \iu$. 
From the above calculation, $F_X \sim \del_X W = \Lambda^2 Y $ can be estimated at the CP-conserving vacuum and is listed in Tab.~\ref{tab:values}.
Our models tend to have the low SUSY breaking scale of $\mc{O}(1)$ TeV due to the suppression by the modular form at the CP-conserving vacuum.

\subsection{Non-gaussianity}

The non-gaussianity in the multi-field inflation is discussed in Refs.~\cite{Lyth:2005fi,Bassett:2005xm,Seery:2005gb}, where the authors would consider the canonically normalized scalar fields.
In the general kinetic term case, this result would be extended to 
\begin{align}
    - \frac{3}{5} f_{\NL} \approx \frac{\msc{N}_{ab} N^{,a} N^{,b}}{2 [K^{ab} N_{,a} N_{,b}]^2}
\end{align}
where we introduce
\begin{align}
	\msc{N}_{ab} \coloneqq N_{,b;a},
\end{align}
so that the covariance of the scalar field space is respected.

\medskip

\begin{figure}[t]
	\centering
	\begin{subfigure}{0.48\textwidth}
		\centering
		\includegraphics[width=\textwidth]{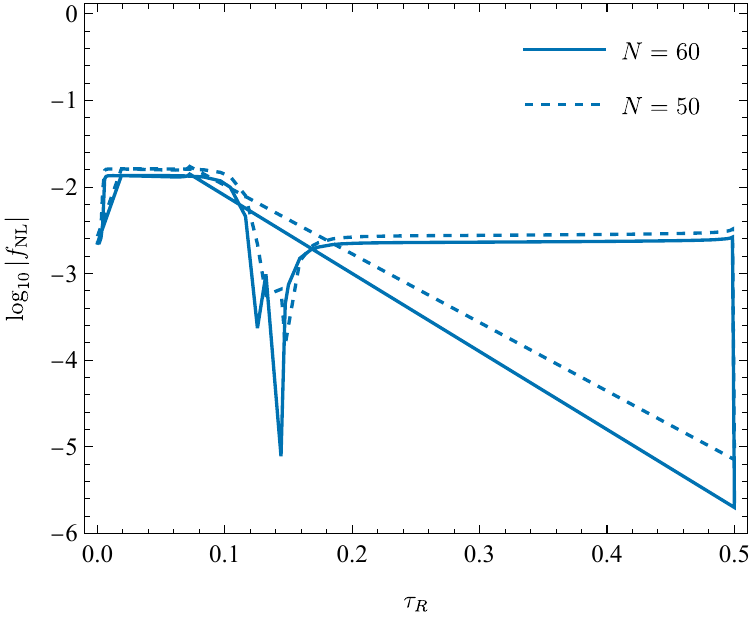}
		\smallskip\\
		\includegraphics[width=\textwidth]{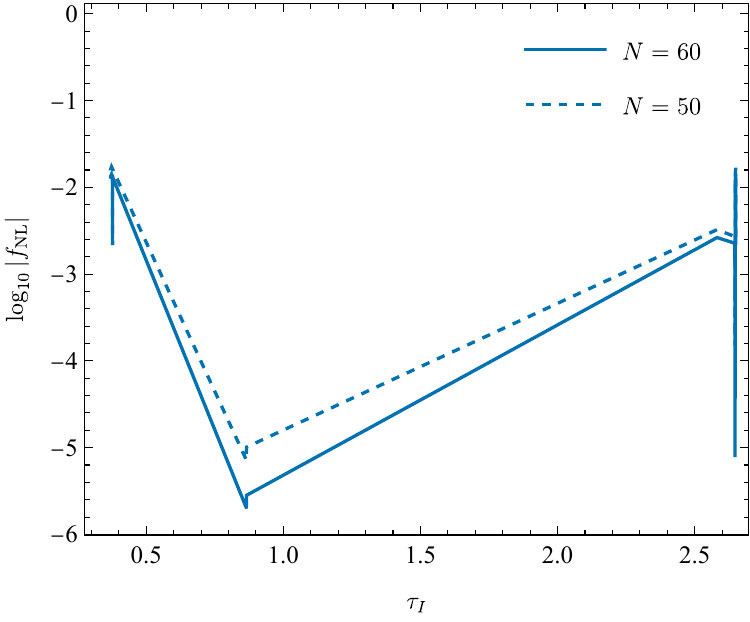}
		\caption{$\alpha = 0.1$}
	\end{subfigure}
	\quad
	\begin{subfigure}{0.48\textwidth}
		\centering
		\includegraphics[width=\textwidth]{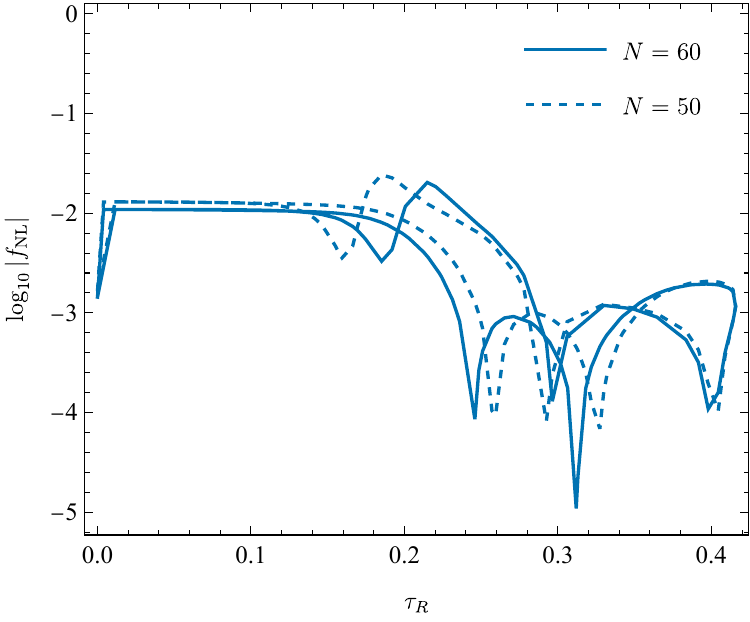}
		\smallskip\\
		\includegraphics[width=\textwidth]{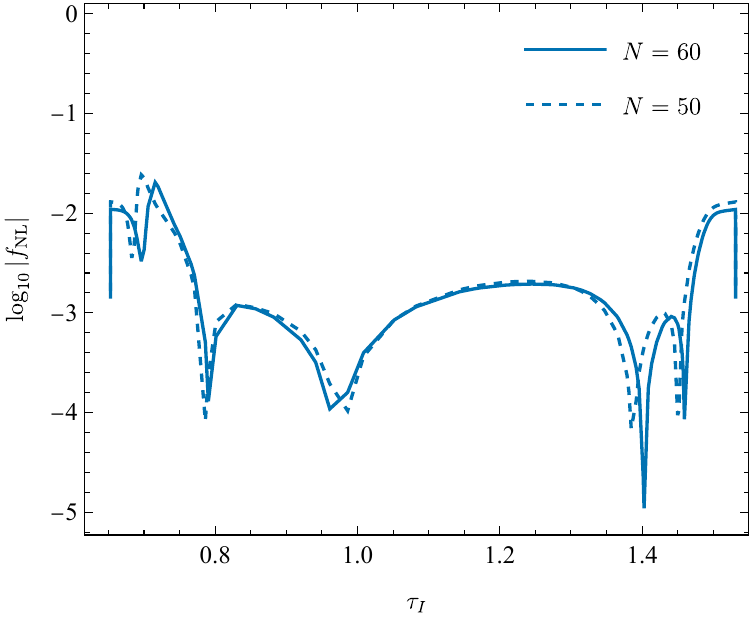}
		\caption{$\alpha = 5$}
	\end{subfigure}
	\caption{The non-gaussianity 
 as the function of the moduli in our model when $\alpha = 0.1$ and 5.
	The blue solid and dashed lines correspond to $f_{\NL}$ at $N=60$ and $N=50$ respectively.}
	\label{fig:model1-taulogfNL}
\end{figure}

From the current observation~\cite{Planck:2019kim}, the absolute value of the non-gaussianity is bounded typically by unity,
\begin{align}
	|f_{\NL}| \lesssim 1.
    \label{eq:constraint-fNL}
\end{align}
We show $f_\NL$ as the function of the moduli at $N=60$ (blue solid lines) and $N=50$ (dashed ones) in our model in Fig.~\ref{fig:model1-taulogfNL} when $\alpha = 0.1$ and 5.
The sharp fins in Fig.~\ref{fig:model1-taulogfNL} correspond to the signature flip of $f_{\NL}$.
$f_\NL$ in our model is consistent with the current constraint.
In our analysis, we focus on the adiabatic perturbation on the inflationary trajectories which can be reduced to the single-field one effectively,
and then a small non-gaussianity given by the slow-roll parameters is consistent with the study in Ref.~\cite{Maldacena:2002vr}.
Analysis of the isocurvature perturbation orthogonal to the adiabatic one is left for the future work though it can give the additional contributions to observations.

\subsection{Inflaton decay}

In this subsection, we discuss decay modes of the moduli after the inflation.
After the inflation, moduli move to the vacuum at $\tau = \iu$, start to oscillate around this vacuum and reheat the Universe via the decay finally. 
We expand the moduli fields as 
\begin{align}
    \tau_R = \delta \tau_R,
    \qquad 
    \tau_I = 1 + \delta \tau_I,
\end{align}
where $\delta \tau_{R,I}$ denote the fluctuations around this vacuum.
The Lagrangian is written as 
\begin{align}
    \bm{L} &= - \frac{h M_P^2}{4} \bigl[
        (\del_\mu \delta \tau_R)^2 + (\del_\mu \delta \tau_I)^2
    \bigr] - V(\delta \tau_R, 1 + \delta \tau_I) 
    \nn \\
    & \quad 
    + \frac{h M_P^2}{2} \delta \tau_I \bigl[
        (\del_\mu \delta \tau_R)^2 + (\del_\mu \delta \tau_I)^2
    \bigr]+ \cdots,
\end{align}
where the last term comes from $1/\tau_I^2 =1/ (1 + \delta \tau_I)^2$ in the non-canonical kinetic term \eqref{eq:kinetic-term-modulus}.
In the canonical kinetic term base,
\begin{align}
    \bm{L} \ni - \frac{1}{2} \bigl[
        (\del_\mu \delta \hat{\tau}_R)^2 + (\del_\mu \delta \hat{\tau}_I)^2
    \bigr],
    \qquad
    \delta \hat{\tau}_{R,I} \coloneqq \sqrt{\frac{h}{2}} M_P \delta \tau_{R,I},
\end{align}
the mass squared matrix of $\tau$ is 
\begin{align}
    M_{\text{modulus}}^2
    \approx 5.67 \times 10^{3} \biggl( \frac{2}{h} \biggr) \biggl(
        \frac{\Lambda}{M_P}
    \biggr)^4 M_P^2 \begin{pmatrix}
        1 & 0 \\
        0 & 1
    \end{pmatrix}.
\end{align}
Here, the mass matrix is given by the second-order derivative of the scalar potential \eqref{eq:model1-potential} in the vacuum and almost insensitive to values of $\alpha$.
The absence of the mixing between $\tau_R$ and $\tau_I$ is the consequence of the CP preserving vacuum.
These two fields have the same mass in the vacuum.
This diagonal terms have a very small difference, but that is able to be ignored in our precision.
$\Lambda/ M_P$ is determined by the PLANCK normalization, and we found $\Lambda/M_P \sim 10^{-(3 \text{--}4)}$ typically as shown in Tab.~\ref{tab:values}.
Using these results, the modulus mass with $h =2$ is roughly given by 
\begin{align}
    m_{\tau} \sim 1.81 \times 10^{12 \text{--} 14}~\mr{GeV}.
\end{align}

\medskip
Suppose that there exists in the action
\begin{align}
    c \int d^2 \theta \bigg(\frac{-\iu \tau}{4\pi}\bigg) 
    {\cal W}^\alpha {\cal W}_\alpha + {\rm h.c.},
\end{align}
where $\theta$ is the fermionic coordinate in the superspace, ${\cal W}^\alpha$ is the superfield strength and correction terms in the gauge coupling which might exist in the modular invariant theories are neglected.
$c$ is a constant associated with this gauge fields, which is treated as the free parameter here.
Such a modulus dependent term would appear in threshold  corrections of gauge kinetic functions in the modular symmetric theory and $c$ may be suppressed by a loop factor \cite{Dixon:1990pc,Kaplunovsky:1994fg,Kaplunovsky:1995jw,Blumenhagen:2006ci}.
Then, in our model the main decay channels will be $\tau_R \to AA$ and $\tau_R \to \psi_{3/2} \psi_{3/2}$, where $A$ and $\psi_{3/2}$ denote the MSSM gauge fields and the gravitino respectively\footnote{
The inflaton can directly decay to the SM particles through the Yukawa coupling
\begin{align}
    \bm{L}_{\delta \tau H \ol{\psi} \psi} \sim - (\del_{\tau} Y) \delta \tau \ol{\psi}^\dagger H \psi,
\end{align}
but the scaling of the decay width is $\Gamma_{\delta \tau \to H \ol{\psi} \psi} \propto m_{\tau}^3/ M_P$ and there is an additional phase suppression factor.
}.
The latter decay mode originates from the gravitino mass term
$\ee^{\mc{K}/2} W \psi_{3/2} \psi_{3/2}$, where 
the gravitino mass is given by $m_{3/2} = \ee^{\mc{K}/2}W$ in the vacuum, and it is expected that $m_\tau \gg m_{3/2} \sim F_X ={\cal O}(1)$ TeV since $V \sim |F_X|^2 - 3m_{3/2}^2 \sim 0$.
The interaction terms read
\begin{align}
    \bm{L}_{\tau F^2} \sim c \delta \tau_R F_{\mu\nu} \tilde{F}^{\mu\nu} + c \delta \tau_I F_{\mu\nu} F^{\mu\nu},
    \qquad 
      \bm{L}_{\tau \psi \psi} \sim \delta \tau F^\tau \psi_{3/2}  \psi_{3/2},
\end{align}
where the dual field strength is introduced by $\tilde{F}^{\mu\nu} \coloneqq \frac{1}{2} \epsilon^{\mu\nu\rho\sigma} F_{\rho\sigma}$
and we have expand the gravitino mass around the vacuum
to read the interaction as $\ee^{\mc{K}/2}W \ni \delta \tau \partial_\tau (\ee^{\mc{K}/2}W) \sim \delta \tau F^\tau$.
We parameterize F-component of $\tau$ as
$F^\tau \sim d (m_{3/2}^2/m_\tau)$ \cite{Nakamura:2006uc,Endo:2006zj} because $F^\tau$ depends on the magnitude of $X$ in the vacuum.
Using these interaction terms, the decay widths of $\tau$ can be estimated as 
\begin{align}
    \Gamma_{\delta \tau \to AA} &\sim N_g \frac{c^2}{4 \pi} \frac{m_{\tau}^3}{M_P^2},
    \\
    \Gamma_{\delta \tau \to \psi_{3/2} \psi_{3/2}} &\sim 
    d^2  \frac{m_{\tau}^3}{96 \pi M_P^2} \sqrt{1 - \frac{4 m_{3/2}^2}{m_{\tau}^2}}.
    \label{eq:decay-width}
\end{align}
Here $N_g= 12$ for the minimal supersymmetric standard model,
and $M_P$ dependence comes from the canonical normalization of the moduli fields.
For $d \ll 1$, the reheating temperature is given by 
\begin{align}
    T_R \approx \bigg(\frac{90}{\pi^2 g_*(T_R)}\bigg)^{1/4} \sqrt{\Gamma_{\delta \tau \to AA} M_{P}}
    \approx  8.9 \times 10^{7}\ \mr{GeV} \, 
    \bigg( \frac{c}{10^{-2}}\bigg)
    \bigg( \frac{m_\tau}{10^{13}{\rm GeV}}\bigg)^{3/2}.
\end{align}
Here we have used $g_*(T_R) = 915/4$.
Because heavy and unstable gravitinos can be produced through the scattering process with particles in the thermal bath at such a high temperature, 
the condition of $m_{3/2} > {\cal O}(1)$ TeV would be required to evade the constraint on the big bang nucleosynthesis destroyed by the abundant gravitino decays \cite{Kawasaki:2008qe}.
On the other hand,
the decay mode into the gravitinos can be dominant for $d \gg 1$.
Then the Universe would be dominated by the gravitino and its decay products including neutralino dark matter.
The big bang nucleosynthesis would be destroyed by the gravitino decay and the Universe would be overclosed. 
Hence reheating by the inflaton decay would then fail \cite{Nakamura:2006uc,Endo:2006zj}. 
This problem could be ameliorated and the baryon asymmetry could be produced via leptogenesis if the gravitino abundance is diluted by a late-time entropy production for $m_{3/2} > {\cal O}(1)$ TeV \cite{Jeong:2012en}.
After all $m_{3/2} > {\cal O}(1)$ TeV in the vacuum is a solution for the gravitino problem, although $F_X \sim {\cal O}(1)$ TeV and therefore $V \sim |F_X|^2 -3m_{3/2}^2$ may become negative.

\medskip
To elude this problem, for instance, $X$ can be considered irrelevant to the SUSY breaking in the vacuum. Another source $X'$ is then supposed to exist and to make the gravitino 
much heavier than ${\cal O}(1)$ TeV in the vacuum; the superpotential relevant to the vacuum can be given by 
$W = \Lambda^2 XY + \mu^2 X'$, where $X'$ is the genuine SUSY breaking field in the vacuum, has the modular weight $-h$, 
and $\mu$ is a constant.
Here, $\mu \ll \Lambda$ but $\mu^2 \gg \Lambda^2 \langle Y \rangle $ in the vacuum;
$\mu^2/M_{P} \sim F_{X'}/M_{P} \sim m_{3/2} \gg F_X \sim  {\cal O}(1)$ TeV.
$X'$ and $\mu$ are considered irrelevant to the inflationary dynamics.

\section{Conclusion}
\label{sec:conclusion}

In this paper, we study the inflation model controlled by the modular flavor symmetry, where the moduli fields play the role not only in driving the inflaton but also in determining the flavor structure.
The extra singlet scalar $X$, namely stabilizer field, is introduced to generate the modulus potential, which is assumed to have a K\"ahler potential in \eqref{eq:model0-Kahlerpotential} and
a simple superpotential~\eqref{eq:superpotential-form}. 
This simple model does not realize 
the slow-roll inflation owing to the steep potential
without the modification of the K\"ahler potential.
To make the scalar potential flatter,
we introduce the additional K\"{a}hler potential~\eqref{eq:model1-deltaK}, which corrects the kinetic term of the stabilizer field $X$ and depends on the modular form included in the superpotential. 
This contribution to the K\"{a}hler potential gives the scalar potential in \eqref{eq:model1-potential} and
can realize the slow-roll inflation successfully.
We show that the parameter space is consistent with the slow-roll inflation and the current observations.
In particular, when the contribution from $\delta K$ becomes larger, the potential becomes much flatter and hence inflaton is given by a combination of not only $\tau_I$ but also $\tau_R$, sharing the same behavior of the $\alpha$-attractor models.
$\tau_I$ can drive the slow-roll inflation around $\tau_R \sim 0$ in terms of the pole inflation in the wide range of $\alpha$ in Eq.~\eqref{eq:model1-deltaK}, whereas
$\tau_R$ can play the role of the waterfall field at the end 
of the hybrid inflation driven by $\tau_I$ for a small $\alpha$.

\medskip
Our analysis in this work focuses on the adiabatic perturbation on the inflationary trajectories, where the isocurvature perturbation is assumed to be small and neglected.
However, the modulus which is the orthogonal to the inflaton direction can be also lighter than the Hubble scale during the inflation. 
This mode can produce the isocurvature perturbation depending on the dynamics of the moduli fields, and would give the additional contribution to the non-gaussianity in our model, which could be tested by future observations.
A more precise analysis of our model is left for the future work.

\medskip
After the inflation, the modulus field rolls down to the CP-symmetric vacuum $\tau = \iu$ at which the inflaton reheats the Universe through the decay of moduli to the gauge fields and the gravitino. 
To generate the baryon asymmetry of the Universe, it will be required to break the CP symmetry at the vacuum in the case of spontaneous CP violation, corresponding to a slight deviation of $\tau$ from $\tau = \iu$. It would be realized by an additional small correction dependent of the moduli to the K\"ahler potential and superpotential, other uplifting mechanisms to obtain the current cosmological constant or the stabilization of K\"{a}hler moduli of the torus from the UV point of view.\footnote{Such a mechanism was discussed in the context of modular flavor models \cite{Ishiguro:2022pde}.} When the CP symmetry is broken by some mechanism or parameters, one of the realistic mechanisms to generate the baryon asymmetry will be the non-thermal leptogenesis via the inflaton decay to the right-handed neutrino.

\medskip
On the CP-symmetric vacuum, there also exists the residual discrete symmetry in the moduli space of $\tau$. 
Since the inflation mechanism is successfully realized by the weight 6 modular form of a finite modular group of $\Gamma_N$, the residual symmetry would play an important role of determining the flavor structure of quarks and leptons for several modular flavor models (e.g., Ref. \cite{Feruglio:2022kea}). 
Furthermore, the $F$-term of stabilizer field $X$ can induce the SUSY breaking at the vacuum in addition to the de Sitter expansion. The typical SUSY breaking scale is ${\cal O}(1)$ TeV due to the suppression by the modular form at the CP-symmetric vacuum $\tau = \iu$. The stabilization of $X$ will be realized by another higher-order term in the K\"{a}hler potential $ \mc{K}\sim - |X|^4$. 
It is interesting to explore the SUSY phenomenology which will be left for future work.

\medskip
In this paper, matter couplings to moduli stabilize them via $q$-dependent corrections with $q = \exp ( 2\pi \iu \tau)$.
The matter contribution might be the interesting source to stabilize the moduli and realize the de Sitter vacua in string theory.

\section*{Acknowledgments}
\noindent
The authors thank to Kohei Kamada and Keigo Shimada for useful comments.
The work of Y.A. is supported by JSPS Overseas Research Fellowships.
This work is supported in part by JSPS Grant-in-Aid for KAKENHI Grant No. JP22K03601 (T.H.), JP20K14477 (H.O.).

\appendix

\section{Modular forms}
\label{app:modular-forms}

In this section, we summarize the modular form of the finite modular group $\Gamma_N$ with $N=3,4,5$. 
In particular, we focus on the singlet modular forms and their $q$ expansions. 
In the manuscript, we deal with the weight 6 modular form of level 3
as the concrete modular form, but it is applicable to other modular groups as shown below.

\subsection{$A_4$}

The level 3 and weight 2 triplet modular form is $Y^{3,2}_{\bm{3}}=(Y_1 (\tau), Y_2(\tau), Y_3(\tau))^\ttr$~\cite{Feruglio:2017spp}, and the components are given by
\begin{align}
	Y_1(\tau) &= \frac{\iu}{2\pi} \biggl(
		\frac{\eta'(\tau/3)}{\eta(\tau/3)} + \frac{\eta'((\tau+1)/3)}{\eta((\tau+1)/3)} + \frac{\eta'((\tau+2)/3)}{\eta((\tau+2)/3)} - 27 \frac{\eta'(3\tau)}{\eta(3\tau)}
	\biggr),
	\\
	Y_2(\tau)&= \frac{-\iu}{\pi} \biggl(
		\frac{\eta'(\tau/3)}{\eta(\tau/3)} + \omega^2 \frac{\eta'((\tau+1)/3)}{\eta((\tau+1)/3)} + \omega \frac{\eta'((\tau+2)/3)}{\eta((\tau+2)/3)}
	\biggr),
	\\
	Y_3(\tau)&= \frac{-\iu}{\pi} \biggl(
		\frac{\eta'(\tau/3)}{\eta(\tau/3)} + \omega \frac{\eta'((\tau+1)/3)}{\eta((\eta+1)/3)} + \omega^2 \frac{\eta'((\tau+2)/3)}{\eta((\tau+2)/3)}
	\biggr),
\end{align}
where $\omega = \ee^{2 \pi \iu /3}$.
They satisfy the following constraint:
\begin{align}
	Y_2^2 + 2 Y_1 Y_3 = 0.
\end{align}
The Dedekind's $\eta$ function is defined by 
\begin{align}
	\eta(\tau) = q^{1/24} \prod_{m=1}^\infty (1 - q^m),
	\qquad 
	q \coloneqq \ee^{2 \pi \iu \tau}.
	\label{eq:eta-def}
\end{align}
Using this definition, $\eta'/\eta$ is written as the following form:
\begin{align}
	\frac{\eta'(\tau)}{\eta(\tau)} =2 \pi \iu \biggl[
		\frac{1}{24} + \sum_{m\geq1} \frac{m}{1 - q^{-m}}
	\biggr],
	\label{eq:eta'/eta}
\end{align}
and, for example, $\eta'((\tau+a)/3)/\eta((\tau+a)/3)$ is explicitly written as 
\begin{align}
    \frac{\eta'((\tau+a)/3)}{\eta((\tau+a)/3)} = 2 \pi \iu \biggl[
        \frac{1}{24} + \sum_{m \geq 1} \frac{m}{1 - q^{-m/3} \ee^{- 2 \pi \iu a m /3}}
    \biggr].
\end{align}

The modular forms with the higher weights are constructed as the products of $Y_i(\tau)$ introduced in the previous subsection.
Here, we will summarize the modular forms with weight 4, 6, and 8.

\paragraph{Weight 4:}
\begin{align}
	Y^{3,4}_{\bm{1}} &= Y_1^2 + 2 Y_2 Y_3,
	\\
	Y^{3,4}_{\bm{1}'} &= Y_3^2 + 2 Y_1 Y_2,
	\\
	Y^{3,4}_{\bm{3}} &= \begin{pmatrix}
		Y_1^2 - Y_2 Y_3 \\
		Y_3^2 - Y_1 Y_2 \\
		Y_2^2 - Y_3 Y_1
	\end{pmatrix}.
\end{align}

\paragraph{Weight 6:}

\begin{align}
	Y^{3,6}_{\bm{1}}&= Y^3_1 + Y^3_2 + Y^3_3 - 3 Y_1 Y_2 Y_3,
	\\
	Y^{3,6}_{\bm{3}} &= Y^{3,2}_{\bm{3}} Y^{3,4}_{\bm{1}} 
	=(Y_1^2 + 2 Y_2 Y_3) \begin{pmatrix}
		Y_1 \\
		Y_2 \\
		Y_3
	\end{pmatrix},
	\\
	Y^{3,6}_{\bm{3}'} &= Y^{3,2}_{\bm{3}} Y^{3,4}_{\bm{1}'} 
	= (Y_3^2 + 2 Y_1 Y_2) \begin{pmatrix}
		Y_3 \\
		Y_1 \\
		Y_2
	\end{pmatrix}.
\end{align}

\paragraph{Weight 8:}

\begin{align}
	Y^{3,8}_{\bm{1}} &=(Y^{3,4}_{\bm{1}})^2
	= (Y^2_1 + 2 Y_2 Y_3)^2,
	\\
	Y^{3,8}_{\bm{1}'} &= Y^{3,4}_{\bm{1}} Y^{3,4}_{\bm{1}'}
	= (Y_1^2+ 2 Y_2 Y_3)(Y_3^2 + 2 Y_1 Y_2),
	\\
	Y^{3,8}_{\bm{1}''}&= (Y^{3,4}_{\bm{1}'})^2 
	= (Y^2_3 + 2 Y_1 Y_2)^2,
	\\
	Y^{3,8}_{\bm{3}} &=(Y_1^2 + 2 Y_2 Y_3) \begin{pmatrix}
		Y_1^2 - Y_2 Y_3 \\
		Y_3^2 - Y_1 Y_2 \\
		Y_2^2 - Y_1 Y_3
	\end{pmatrix},
	\\
	Y^{3,8}_{\bm{3}'} &= ( Y_3^2 + 2 Y_1 Y_2) \begin{pmatrix}
		Y_2^2 - Y_1 Y_3 \\
		Y_1^2 - Y_2 Y_3 \\
		Y_3^2 - Y_1 Y_2
	\end{pmatrix}.
\end{align}

If $\ee^{- 2 \pi \tau_I} \ll 1$, we can consider the expansion of $q$.
The components of $Y^{3,2}_{\bm{3}}$ are written as 
\begin{align}
    Y_1 &= 1 + 12 q + 36 q^2 + 12 q^3 + 84 q^4 + \mc{O}(q^{13/3}),
    \\
    Y_2 &= - 6 q^{1/3} ( 1 + 7 q + 8 q^2 + 18 q^3) + \mc{O}(q^{13/3}),
    \\
    Y_3 &= - 18 q^{3/2}( 1 + 2 q + 5 q^2 + 4 q^3) + \mc{O}(q^{13/3}).
\end{align}
In the same manner, the singlets are expressed as 
\begin{align}
    Y^{3,4}_{\bm{1}} &= 1 + 240 q + 2160 q^2 + 6720 q^3 + 17520 q^4 + \mc{O}(q^{13/3}),
    \label{eq:Y^{3,4}_1-q-exp}
    \\
    Y^{3,6}_{\bm{1}} &= 1 - 504 q - 16632 q^2 - 122976 q^3 - 532728 q^4 + \mc{O}(q^{13/3}).
    \label{eq:Y^{3,6}_1-q-exp}
\end{align}
Note that the above singlets are also described by Eisenstein series $E_4$ and $E_6$, respectively.

\subsection{$S_4$}

The weight 2 modular form of $\Gamma_4 \simeq S_4$ was constructed in Ref. \cite{Penedo:2018nmg}. 
We first define 
\begin{align}
    Y_1^{4}(\tau) &= Y(1,1,\omega, \omega^2, \omega, \omega^2,\tau),
    \nonumber\\
    Y_2^{4}(\tau) &= Y(1,1,\omega^2, \omega, \omega^2, \omega,\tau),
    \nonumber\\
    Y_3^{4}(\tau) &= Y(1,-1,-1,-1,1,1,\tau),
    \nonumber\\
    Y_4^{4}(\tau) &= Y(1,-1,-\omega^2, -\omega, \omega^2, \omega,\tau),
    \nonumber\\
    Y_5^{4}(\tau) &= Y(1,-1,-\omega, -\omega^2, \omega, \omega^2,\tau),
\end{align}
with 
\begin{align}
    Y(a_1, a_2, a_3, a_4, a_5, a_6, \tau) = a_1 \frac{\eta'(\tau+1/2)}{\eta(\tau+1/2)} +4a_2 \frac{\eta'(4\tau)}{\eta(4\tau)} + \frac{1}{4} \sum_{m=0}^3 a_{m+3}\frac{\eta'((\tau+m)/4)}{\eta((\tau+m)/4)}.
\end{align}
The weight 2 modular forms of the level 4 is of the form:
\begin{align}
	Y^{4,2}_{\bm{2}} &=\begin{pmatrix}
		Y_1^4(\tau) \\
            Y_2^4(\tau)
	\end{pmatrix},
	\quad
	Y^{4,2}_{\bm{3}'} = \begin{pmatrix}
		Y_3^4(\tau) \\
            Y_4^4(\tau) \\
            Y_5^4(\tau) \\
	\end{pmatrix}.
\end{align}

The modular form with the higher weights are constructed by the tensor product of the weight 2 
modular forms. In the following, we will summarize the modular forms with weight 4, 6, and 8 \cite{Novichkov:2018ovf}.

\paragraph{Weight 4:}
\begin{align}
\begin{split}
	Y^{4,4}_{\bm{1}} &= Y_1^4 Y_2^4,
	\quad
	Y^{4,4}_{\bm{2}} = \begin{pmatrix}
		(Y_2^{4})^2\\
		(Y_1^{4})^2
	\end{pmatrix},
	\\
	Y^{4,4}_{\bm{3}} &= \begin{pmatrix}
		Y_1^{4}Y_4^{4} - Y_2^{4}Y_5^{4}\\
		Y_1^{4}Y_5^{4} - Y_2^{4}Y_3^{4}\\
		Y_1^{4}Y_3^{4} - Y_2^{4}Y_4^{4}
         \end{pmatrix},
	\quad
	Y^{4,4}_{\bm{3}'} = \begin{pmatrix}
		Y_1^{4}Y_4^{4} + Y_2^{4}Y_5^{4}\\
		Y_1^{4}Y_5^{4} + Y_2^{4}Y_3^{4}\\
		Y_1^{4}Y_3^{4} + Y_2^{4}Y_4^{4}
 \end{pmatrix}.
\end{split}
\end{align}

\paragraph{Weight 6:}

\begin{align}
\begin{split}
	Y^{4,6}_{\bm{1}}&= (Y^4_1)^3 + (Y^4_2)^3,
	\quad
	Y^{4,6}_{\bm{1}'}= (Y^4_1)^3 - (Y^4_2)^3,
	\\
	Y^{4,6}_{\bm{2}} &= Y^4_1 Y^4_2\begin{pmatrix}
		Y_1^{4}\\
		Y_2^{4}
	\end{pmatrix},
         	\quad
	Y^{4,6}_{\bm{3}} = \begin{pmatrix}
		(Y_2^{4})^2 Y_4^{4} - (Y_1^{4})^2 Y_5^{4}\\
		(Y_2^{4})^2 Y_5^{4} - (Y_1^{4})^2 Y_3^{4}\\
		(Y_2^{4})^2 Y_3^{4} - (Y_1^{4})^2 Y_4^{4}
         \end{pmatrix},
         \\
	Y^{4,6}_{\bm{3}',I} &= Y^4_1 Y^4_2\begin{pmatrix}
		Y_3^{4}\\
		Y_4^{4}\\
		Y_5^{4}\\
         \end{pmatrix},
         	\quad
	Y^{4,6}_{\bm{3}',II} = \begin{pmatrix}
		(Y_2^{4})^2 Y_4^{4} + (Y_1^{4})^2 Y_5^{4}\\
		(Y_2^{4})^2 Y_5^{4} + (Y_1^{4})^2 Y_3^{4}\\
		(Y_2^{4})^2 Y_3^{4} + (Y_1^{4})^2 Y_4^{4}
         \end{pmatrix}.
\end{split}
\end{align}

\paragraph{Weight 8:}

\begin{align}
\begin{split}
        Y^{4,8}_{\bm{1}}&= (Y^4_1)^2(Y^4_2)^2,
	\quad
	Y^{4,8}_{\bm{2},I} = Y^4_1Y^4_2\begin{pmatrix}
		(Y_2^{4})^2\\
		(Y_1^{4})^2
	\end{pmatrix},
	\quad
	Y^{4,8}_{\bm{2},II} = ((Y^4_1)^3 - (Y^4_2)^3)\begin{pmatrix}
		Y_1^{4}\\
		-Y_2^{4}
	\end{pmatrix},
        \\
	Y^{4,8}_{\bm{3},I} &= ((Y^4_1)^3 - (Y^4_2)^3)\begin{pmatrix}
		Y_3^{4}\\
		Y_4^{4}\\
		Y_5^{4}\\
         \end{pmatrix},
        \quad
	Y^{4,8}_{\bm{3},II} = Y^4_1 Y^4_2\begin{pmatrix}
		Y_1^{4}Y_4^{4} - Y_2^{4}Y_5^{4}\\
		Y_1^{4}Y_5^{4} - Y_2^{4}Y_3^{4}\\
		Y_1^{4}Y_3^{4} - Y_2^{4}Y_4^{4}
         \end{pmatrix},
         	\\
	Y^{4,8}_{\bm{3}',I} &= ((Y^4_1)^3 + (Y^4_2)^3)\begin{pmatrix}
		Y_3^{4}\\
		Y_4^{4}\\
		Y_5^{4}\\
         \end{pmatrix},
        \quad
        Y^{4,8}_{\bm{3}',II} = Y^4_1 Y^4_2\begin{pmatrix}
		Y_1^{4}Y_4^{4} + Y_2^{4}Y_5^{4}\\
		Y_1^{4}Y_5^{4} + Y_2^{4}Y_3^{4}\\
		Y_1^{4}Y_3^{4} + Y_2^{4}Y_4^{4}
         \end{pmatrix}.
\end{split}
\end{align}

Let us consider the $q$ expansion $\ee^{- 2 \pi \tau_I} \ll 1$. 
Since $Y_1^4$ and $Y_2^4$ are expanded as
\begin{align}
    Y_1^4 &= \frac{3\pi \iu}{8}\left( 1 - 8 \iu \sqrt{3q} +24 q -32 \iu \sqrt{3} q^{3/2} + 24 q^2 -\frac{64 \iu}{\sqrt{3}}q^{5/2} + 64q^3 +\mc{O}(q^{7/2})\right),
    \\
    Y_2^4 &= \frac{3\pi \iu}{8}\left( 1 + 8 \iu \sqrt{3q} +24 q + 32 \iu \sqrt{3} q^{3/2} + 24 q^2 +\frac{64 \iu}{\sqrt{3}}q^{5/2} + 64q^3 +\mc{O}(q^{7/2})\right),
\end{align}
the trivial singlets are expressed as
\begin{align}
    Y^{4,4}_{\bm{1}} &= -\frac{9\pi^2}{64}\left( 1 + 240 q + 2160 q^2 + 6720 q^3 + 17520 q^4 + \mc{O}(q^{5})\right),
    \\
    Y^{4,6}_{\bm{1}} &= -\frac{27\pi^3 \iu}{256}\left(1 - 504 q - 16632 q^2 - 122976 q^3 - 532728  q^4 + \mc{O}(q^{5})\right),
    \\
    Y^{4,8}_{\bm{1}} &= \frac{81\pi^4}{4096}\left(1 + 480 q + 61920 q^2 +1050240 q^3 + 7926240 q^4 + \mc{O}(q^{5})\right).
\end{align}
Thus, these modular forms of level 4 are the same expansion as those of level 3 up to the overall factor. 
Note that the singlet $Y^{4,8}_{\bm{1}}$ is also described by the Eisenstein series $E_8$ in the same manner as the other singlets.

\subsection{$A_5$}

The dimension of weight 2 modular forms of $\Gamma_5 \simeq A_5$ is 11. 
Following the notation of Ref. \cite{Novichkov:2018nkm}, we define the weight 2 modular forms of level 5:
\begin{align}
    Y_1^{5}(\tau) &= -\frac{1}{\sqrt{6}}Y^{(5)}(-5,1,1,1,1,1; -5,1,1,1,1,1|\tau),
    \nonumber\\
    Y_2^{5}(\tau) &= Y^{(5)}(0, 1, \zeta^4, \zeta^3, \zeta^2, \zeta; 0, 1, \zeta^4, \zeta^3, \zeta^2, \zeta|\tau),
    \nonumber\\
    Y_3^{5}(\tau) &= Y^{(5)}(0, 1, \zeta^3, \zeta, \zeta^4, \zeta^2; 0, 1, \zeta^3, \zeta, \zeta^4, \zeta^2|\tau),
    \nonumber\\
    Y_4^{5}(\tau) &= Y^{(5)}(0, 1, \zeta^2, \zeta^4, \zeta, \zeta^3; 0, 1, \zeta^2, \zeta^4, \zeta, \zeta^3|\tau),
    \nonumber\\
    Y_5^{5}(\tau) &= Y^{(5)}(0, 1, \zeta, \zeta^2, \zeta^3, \zeta^4; 0, 1, \zeta, \zeta^2, \zeta^3, \zeta^4|\tau),
    \nonumber\\
    Y_6^{5}(\tau) &= \frac{1}{\sqrt{2}}Y^{(6)}(-\sqrt{5},-1,-1,-1,-1,-1; \sqrt{5},1,1,1,1,1|\tau),
    \nonumber\\
    Y_7^{5}(\tau) &= Y^{(5)}(0, 1, \zeta^4, \zeta^3, \zeta^2, \zeta; 0, -1, -\zeta^4, -\zeta^3, -\zeta^2, -\zeta|\tau),
    \nonumber\\
    Y_8^{5}(\tau) &= Y^{(5)}(0, 1, \zeta, \zeta^2, \zeta^3, \zeta^4; 0, -1, -\zeta, -\zeta^2, -\zeta^3, -\zeta^4|\tau),
    \nonumber\\
    Y_9^{5}(\tau) &= \frac{1}{\sqrt{2}}Y^{(5)}(\sqrt{5},-1,-1,-1,-1,-1; -\sqrt{5},1,1,1,1,1|\tau),
    \nonumber\\
    Y_{10}^{5}(\tau) &= Y^{(5)}(0, 1, \zeta^3, \zeta, \zeta^4, \zeta^2; 0, -1, -\zeta^3, -\zeta, -\zeta^4, -\zeta^2|\tau),
    \nonumber\\
    Y_{11}^{5}(\tau) &= Y^{(5)}(0, 1, \zeta^2, \zeta^4, \zeta, \zeta^3; 0, -1,  -\zeta^2, -\zeta^4, -\zeta, -\zeta^3|\tau),
\end{align}
where the function $Y^{(5)}$ is defined as
\begin{align}
    Y^{(5)}(c_{1,-1}, c_{1,0},...,c_{1,4}; c_{2,-1}, c_{2,0},...,c_{2,4}| \tau) := \sum_{i,j}c_{i,j} \frac{\dd}{\dd \tau} \log \alpha_{i,j}(\tau)
\end{align}
with $\sum_{i,j}c_{i,j}=0$ and  
\begin{align}
    \begin{split}
        \alpha_{1,-1}(\tau) &= \theta_3\left( \frac{\tau+1}{2}, 5\tau\right),\qquad
        \alpha_{2,-1}(\tau) = e^{2\pi \iu \tau/5} \theta_3\left( \frac{3\tau+1}{2}, 5\tau\right),
        \\
        \alpha_{1,0}(\tau) &= \theta_3\left( \frac{\tau+9}{10}, \frac{\tau}{5}\right),\qquad
        \alpha_{2,0}(\tau) =  \theta_3\left( \frac{\tau+7}{10}, \frac{\tau}{5}\right),
        \\
        \alpha_{1,1}(\tau) &= \theta_3\left( \frac{\tau}{10}, \frac{\tau+1}{5}\right),\qquad
        \alpha_{2,1}(\tau) =  \theta_3\left( \frac{\tau+8}{10}, \frac{\tau+1}{5}\right),
        \\
        \alpha_{1,2}(\tau) &= \theta_3\left( \frac{\tau+1}{10}, \frac{\tau+2}{5}\right),\qquad
        \alpha_{2,2}(\tau) =  \theta_3\left( \frac{\tau+9}{10}, \frac{\tau+2}{5}\right),
        \\
        \alpha_{1,3}(\tau) &= \theta_3\left( \frac{\tau+2}{10}, \frac{\tau+3}{5}\right),\qquad
        \alpha_{2,3}(\tau) =  \theta_3\left( \frac{\tau}{10}, \frac{\tau+3}{5}\right),
        \\
        \alpha_{1,4}(\tau) &= \theta_3\left( \frac{\tau+3}{10}, \frac{\tau+4}{5}\right),\qquad
        \alpha_{2,4}(\tau) =  \theta_3\left( \frac{\tau+1}{10}, \frac{\tau+4}{5}\right).
    \end{split}
\end{align}
Here, $\theta_3(z(\tau), t(\tau))$ is the Jacobi theta function.
The 11-dimensional space of weight 2 modular forms of level 5 is divided into 
\begin{align}
	Y^{2,5}_{\bm{5}} &= \begin{pmatrix}
		Y_1^5 (\tau)\\
		Y_2^5 (\tau)\\
		Y_3^5 (\tau)\\
		Y_4^5 (\tau)\\
		Y_5^5 (\tau)\\
 \end{pmatrix},
 \quad
	Y^{2,5}_{\bm{3}} = \begin{pmatrix}
		Y_6^5 (\tau)\\
		Y_7^5 (\tau)\\
		Y_8^5 (\tau)\\
 \end{pmatrix},
 \quad
	Y^{2,5}_{\bm{3}'} = \begin{pmatrix}
		Y_9^5 (\tau)\\
		Y_{10}^5 (\tau)\\
		Y_{11}^5 (\tau)\\
 \end{pmatrix}.
\end{align}
The higher weight modular forms can be constructed by the tensor product of the above modular forms. 
In the following, we show only the trivial singlet modular forms with weight $k$ of level 5, i.e., $Y_{\bm 1}^{k,5}$ with $k=4,6,8$:
\begin{align}
\begin{split}
    Y^{4,5}_{\bm{1}} &= (Y_1^5)^2 +2Y_3^5Y_4^5 + 2Y_2^5Y_5^5,
    \\
    Y^{6,5}_{\bm{1}} &= 3\sqrt{3}\left( Y_2^5(Y_3^5)^2 +Y_5^5(Y_4^5)^2\right)
    + \sqrt{2}Y_1^5 \left( (Y_1^5)^2 +3Y_3^5Y_4^5 - 6Y_2^5Y_5^5\right),
    \\
    Y^{8,5}_{\bm{1}} &= \left( (Y_1^5)^2 +2Y_3^5Y_4^5 + 2Y_2^5Y_5^5\right) (Y_6^5)^2
    + 2\left( (Y_1^5)^2 +2Y_3^5Y_4^5 + 2Y_2^5Y_5^5\right) Y_7^5Y_8^5.
\end{split}
\end{align}

Let us consider the $q$ expansion $\ee^{- 2 \pi \tau_I} \ll 1$. 
Since $Y_{1,2,3,4,5}^4$ is expanded as
\begin{align}
\begin{split}
    Y_1^5 &= \pi \iu \sqrt{\frac{2}{3}}\left( 1 + 6q     + 18q^2 + 24q^3 + 42 q^4 + 6q^5 +\mc{O}(q^{6})\right),
    \\
    Y_2^5 &= -2\pi \iu \left( q^{1/5} + 12 q^{6/5} + 12 q^{11/5} + 31 q^{16/5} + 32 q^{21/5} +\mc{O}(q^{26/5})\right),
    \\
    Y_3^5 &= -2\pi \iu \left( 3q^{2/5} + 8 q^{7/5} + 28 q^{12/5} + 18 q^{17/5} + 36 q^{22/5} +\mc{O}(q^{27/5})\right),
    \\
    Y_4^5 &= -2\pi \iu \left( 4q^{3/5} + 15 q^{8/5} + 14 q^{13/5} + 39 q^{18/5} + 24 q^{23/5} +\mc{O}(q^{28/5})\right),
    \\
    Y_5^5 &= -2\pi \iu \left( 7q^{4/5} + 13 q^{9/5} + 24 q^{14/5} + 20 q^{19/5} + 60 q^{24/5} +\mc{O}(q^{29/5})\right),
    \\
    Y_6^5 &= -\pi \iu \sqrt{\frac{2}{5}}\left( -1 + 30q     + 20q^2 + 40q^3 + 90 q^4 + 130q^5 +\mc{O}(q^{6})\right),
    \\
    Y_7^5 &= 2\sqrt{5}\pi \iu \left(  q^{1/5} + 2 q^{6/5} + 12 q^{11/5} + 11 q^{16/5} + 12 q^{21/5} +\mc{O}(q^{26/5})\right),
    \\
    Y_8^5 &= 2\sqrt{5}\pi \iu \left(  3q^{4/5} + 7 q^{9/5} + 6 q^{14/5} + 20 q^{19/5} + 10 q^{24/5} +\mc{O}(q^{29/5})\right),
\end{split}
\end{align}
the trivial singlets are expressed as
\begin{align}
    Y^{4,5}_{\bm{1}} &= -\frac{2\pi^2}{3}\left( 1 + 240 q + 2160 q^2 + 6720 q^3 + 17520 q^4 + \mc{O}(q^{5})\right),
    \\
    Y^{6,5}_{\bm{1}} &= -\frac{4\pi^3i}{3\sqrt{3}}\left(1 - 504 q - 16632 q^2 - 122976 q^3 - 532728  q^4 + \mc{O}(q^{5})\right),
    \\
    Y^{8,5}_{\bm{1}} &= \frac{4\pi^4}{15}\left(1 + 480 q + 61920 q^2 +1050240 q^3 + 7926240 q^4 + \mc{O}(q^{5})\right).
\end{align}
Thus, these modular forms of level 5 are the same expansion as those of level 3 and 4.

\section{Multi-field inflation}
\label{app:multi-field-inflation}

In this section, we summarize the results of the multi-field inflation according to Refs.~\cite{Sasaki:1995aw,Chiba:2008rp,Salvio:2021lka}.
The action we consider is
\begin{align}
	S = \int \dd^4x \sqrt{-g} \biggl[
		\frac{M_P^2}{2} \mc{R} - \frac{1}{2} K_{ab}(\phi) \del_\mu \phi^a \del_\nu \phi^b g^{\mu\nu} - V(\phi)
	\biggr],
\end{align}
where $K_{ab}(\phi)$ is the metric of the scalar field space and $V(\phi)$ is the scalar potential.
$\mc{R}$ is the Ricci scalar.
In order to discuss the inflationary expansion, we consider the following configurations:
\begin{align}
	\dd s^2 = - \dd t^2 + a(t)^2 \dd \vec{x}^2,
	\qquad 
	\phi^a(t, \vec{x}) = \phi^a(t),
\end{align}
and the Hubble parameter is defined by $H \coloneqq \dot{a}/ a$.
Dot denotes the time derivative.
The Klein-Gordon equation and Friedmann equations are given by 
\begin{align}
	&\ddot{\phi}^a + 3 H \dot{\phi}^a + \gamma^a_{bc} \dot{\phi}^b \dot{\phi}^c + K^{ab} V_{,b} = 0,
	\qquad 
	H^2 = \frac{1}{3 M_P^2} \biggl(
		\frac{1}{2} K_{ab} \dot{\phi}^a \dot{\phi}^b + V(\phi)
	\biggr), 
    \label{eq:EOMs-KG-Friedmann}
\end{align}
where $V_{,a} \coloneqq \del_a V$.
$\gamma^a_{bc}$ is the connection in the scalar field space, and we introduce the covariant derivative on the scalar field space $\msc{D}_a$ by using this connection:
\begin{align}
    \msc{X}^b_{c;a} \coloneqq \msc{D}_a \msc{X}^b_c = \del_a \msc{X}^b_c + \gamma^b_{ad} \msc{X}^d_c - \gamma^d_{ac} \msc{X}^b_d.
\end{align}
For the metric \eqref{eq:Kahler-metric-tau}, we find that the non-vanishing components of this connection are
\begin{align}
    \gamma^R_{RI} = \gamma^R_{IR} = - \frac{1}{\tau_I},
    \qquad 
    \gamma^I_{RR} = \frac{1}{\tau_I},
    \qquad  
    \gamma^I_{II} = - \frac{1}{\tau_I}.
\end{align}
Here, we use the indices $a, b ,\ldots = R, I$ for the $\tau_R$ and $\tau_I$ components.
The indices are raised and lowered by the metric $K_{ab}$ and $K^{ab}$.
In addition, the time derivative of the Hubble parameter is the scalar field kinetic energy 
\begin{align}
	\dot{H} = - \frac{1}{2 M_P^2} K_{ab} \dot{\phi}^a \dot{\phi}^b.
\end{align}
In the slow-roll regime, where $\epsilon_V <1,~|\eta_V{}^a_b|<1$, these equations reduce to 
\begin{align}
	\dot{\phi}^a \approx - \frac{V^{,a}}{3H},
	\qquad
	H^2 \approx \frac{V}{3 M_P^2},
    \label{eq:EOMs-slow-roll}
\end{align}
where $V^{,a} = K^{ab} V_{,b}$.
With the Friedmann equation, the time derivative of the scalar field is given in terms of the scalar fields as 
\begin{align}
	\dot{\phi}^a \approx - \frac{M_P V^{,a}}{\sqrt{3 V}}.
    \label{eq:dotphi^a-slow-roll}
\end{align}

\subsection{Slow-roll parameters, e-folding, and observables}

The slow-roll parameters are extended by taking the multi-field contributions into the account as 
\begin{align}
	\epsilon_V &\coloneqq \frac{M_P^2}{2} \frac{V_{,a} V^{,a}}{V^2},
	\\
	\eta_V{}^a_b &\coloneqq \frac{M_P^2 V^{;a}_{;b}}{V}.
\end{align}
The end of the inflation is characterized by $\epsilon_V = 1$, and the slow-roll regime is given by $\epsilon_V<1$ and $|\eta_V{}^a_b| < 1$.

\medskip
The e-folding before the end of the inflation is defined by $N = \log (a_f/a)$, where $a_f$ denotes the scale factor at the end of the inflation.
$a$ in the denominator gives the $t$ dependence of $N$.
This is written by the integration of the Hubble parameter as
\begin{align}
	N = \int_{t(\phi)}^{t_f} \dd t' \, H(t'), 
\end{align}
and the derivative of $N$ with respect of $t$ satisfies
\begin{align}
    \frac{\dd N}{\dd t} = -H.
    \label{eq:dNdt=-H}
\end{align}
From this equation, the following useful relation is obtained
\begin{align}
    H = - N_{,a} \dot{\phi}^a \approx N_{,a} \frac{V^{,a}}{3 H},
\end{align}
where the slow-roll EOM \eqref{eq:EOMs-slow-roll} is used in the second equality.
This equation can be formally solved, and $N_{,a}$ is written as 
\begin{align}
    N_{,a} = \frac{1}{M_P^2} \frac{V V_{,a}}{V_{,b} V^{,b}} + \perp_a,
\end{align}
where $\perp_a$ denotes a term orthogonal to $V_{,a}$.
In this work, we focus on the first term.
In the canonically normalized single-field case, this equation reduces to the well-known form 
\begin{align}
    N_\phi = \frac{1}{M_P^2} \frac{V}{\del_\phi V},
\end{align}
because the $\del_\phi V$ in the numerator and denominator are cancelled.

\medskip
The power spectrum $\mc{P}_{\mc{R}}$, spectral index $n_s$, and tensor-to-scalar ratio $r$ are given by $N_{,a}$ and slow-roll parameters as 
\begin{align}
	\mc{P}_{\mc{R}} &= \biggl( \frac{H}{2\pi} \biggr)^2 N_{,a} N^{,a},
	\\
	n_s &= 1 - 2 \epsilon_V - \frac{2}{M_P^2 N_{,a} N^{,a}} + \frac{2 \eta_{Vab}N^{,a} N^{,b}}{N_{,c} N^{,c}},
	\\
	r &\coloneqq \frac{\mc{P}_t}{\mc{P}_{\mc{R}}} = \frac{8}{M_P^2 N_{,a} N^{,a}}.
\end{align}

\subsection{Field equations}
\label{app:field-equations}

In this section, let us rewrite the Klein-Gordon equation in \eqref{eq:EOMs-KG-Friedmann}.
It is useful to use the e-folding $N$ instead of the physical time.
Using \eqref{eq:dNdt=-H}, Eq.~\eqref{eq:EOMs-KG-Friedmann} becomes
\begin{align}
	\phi^{a\prime\prime} - 3 \biggl(
		1 - \frac{1}{6 M_P^2} K_{bc} \phi^{b\prime} \phi^{c\prime}
	\biggr) \biggl(
		\phi^{a\prime} - \frac{M_P^2 K^{ab} V_{,b}}{V}
	\biggr) + \gamma^a_{bc} \phi^{b\prime} \phi^{c\prime} = 0.
    \label{eq:EOM-2-phi}
\end{align}
The prime denotes the derivative with respect to $N$.
When we derive this equation, we use the Friedmann equation, which is written in this case as 
\begin{align}
    H^2 = \frac{\frac{V}{3 M_P^2}}{1 - \frac{1}{6 M_P^2} K_{ab} \phi^{a\prime} \phi^{b\prime}}.
\end{align}
In the slow-roll region, $\phi^{\prime \prime}$ and $(\phi^{\prime})^2$ are dropped and Eq.~\eqref{eq:EOM-2-phi} becomes  
\begin{align}
	\phi^{a\prime} = \frac{M_P^2 V^{,a}}{V},
	\label{eq:phi-prime-EOM}
\end{align}
which is consistent with Eq.~\eqref{eq:dotphi^a-slow-roll}.

\paragraph*{Comments on EOMs after slow-roll.}

\begin{figure}[t]
	\centering
	\begin{subfigure}{0.48\textwidth}
		\centering
		\includegraphics[width=\textwidth]{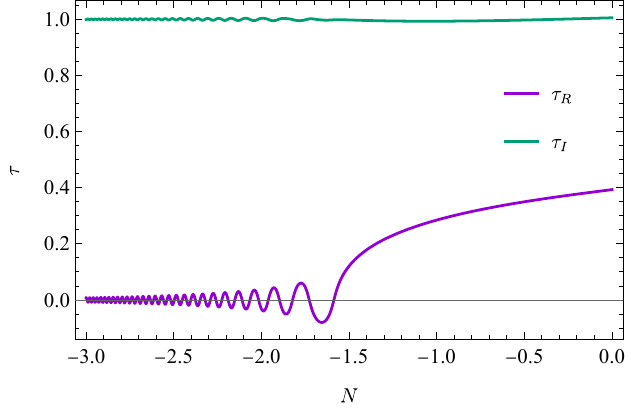}
		\caption{$\alpha = 0.1$}
	\end{subfigure}
	\quad
	\begin{subfigure}{0.48\textwidth}
		\centering
		\includegraphics[width=\textwidth]{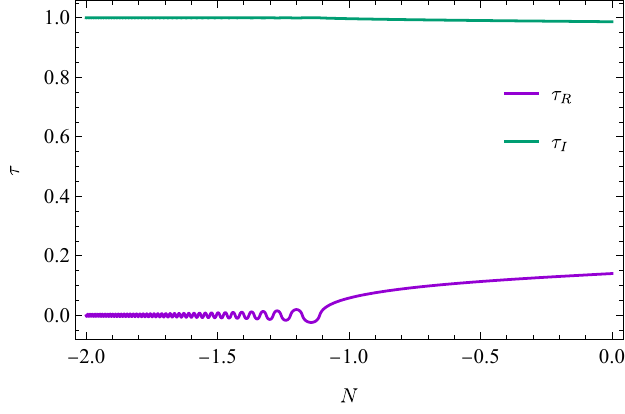}
		\caption{$\alpha = 5$}
	\end{subfigure}
	\caption{The time evolution with $N$ after the slow-roll inflation for $\alpha =0.1$(left) and $5$ (right).
	Moduli fields settle down into the vacuum immediately, oscillating around the vacuum. Note that $N$ formally becomes negative for $|\eta_V{}^a_b|>1$.
 }
	\label{fig:model1-Ntau-end}
\end{figure}

In our model introduced in Sec.~\ref{sec:model}, the EOMs of modulus field \eqref{eq:EOM-2-phi} is given by 
\begin{align}
	&\tau_R'' - 3 \biggl( 1 - h \frac{\tau_R'^2 + \tau_I'^2}{12 \tau_I^2} \biggr)
	\biggl( \tau'_R - \frac{2 \tau_I^2 V_{,\tau_R}}{h V} \biggr) - \frac{2 \tau_R' \tau_I'}{\tau_I} = 0,
	\\
	&\tau_I'' - 3 \biggl( 1 - h \frac{\tau_R'^2 + \tau_I'^2}{12 \tau_I^2} \biggr)
	\biggl( \tau'_I - \frac{2 \tau_I^2 V_{,\tau_I}}{h V} \biggr) + \frac{\tau_R'^2 - \tau_I'^2}{\tau_I} =0.
\end{align}
When moduli cease slow-roll,
we have to use these equations to study the scalar field dynamics after the slow-roll inflation.
The time evolution of the moduli fields after the slow-roll inflation is shown in Fig.~\ref{fig:model1-Ntau-end}.
These time evolutions correspond to the black solid lines at the end of the inflation in Fig.~\ref{fig:model1-vectraj-alp}.
We set $N=0$ at $|\eta_V{}^a_b|=1$ and hence note that $N$ formally becomes negative for $|\eta_V{}^a_b|>1$ after the slow-roll inflation.
Moduli settle down into the vacuum immediately after the end of the slow-roll inflation, oscillating around the vacuum.
From these observations, we use the slow-roll approximation of Eq.~\eqref{eq:phi-prime-EOM} to study the slow-roll inflaton.

\section{Inflation via balance between two matter contributions}
\label{app:Y^{3,6}-and-Y^{3,4}}

\begin{figure}[t]
	\centering
	\includegraphics[width=0.48\textwidth]{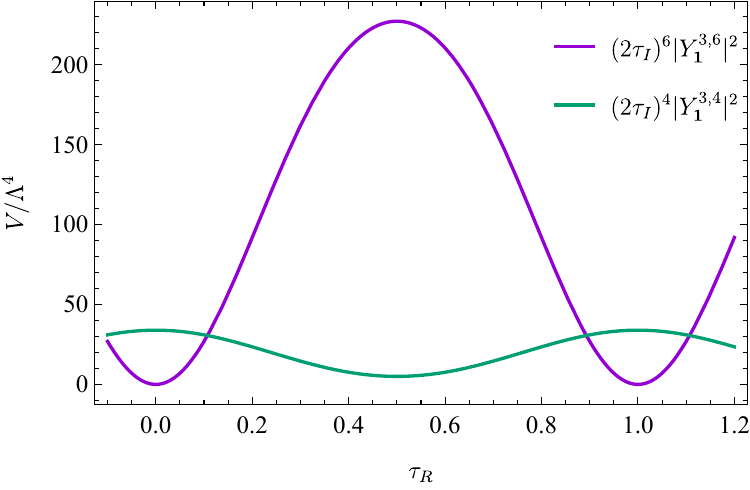}
	\quad
	\includegraphics[width=0.48\textwidth]{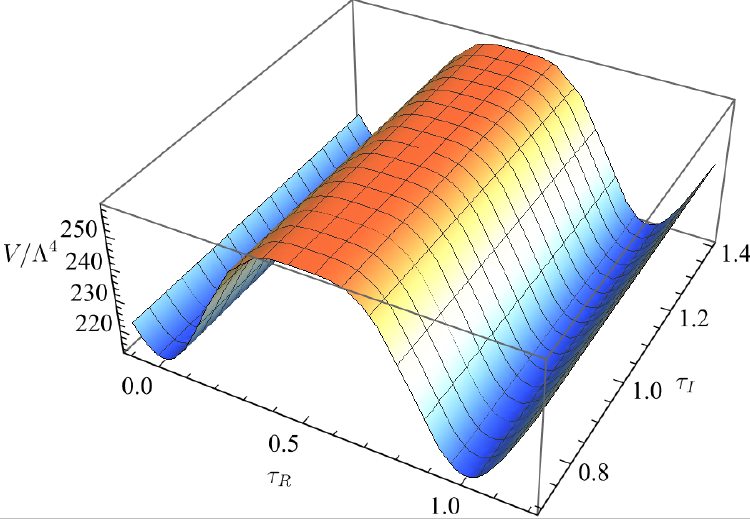}
	\caption{The scalar potential with the suprepotenial~\eqref{eq:model2-superpotential} and the K\"{a}hler potential~\eqref{eq:model2-Kahler-potential}.
	We choose $\beta^2 = 6.2972$.
	}
	\label{fig:model2-potential}
\end{figure}

\begin{figure}[t]
	\centering
	\includegraphics[width=0.48\textwidth]{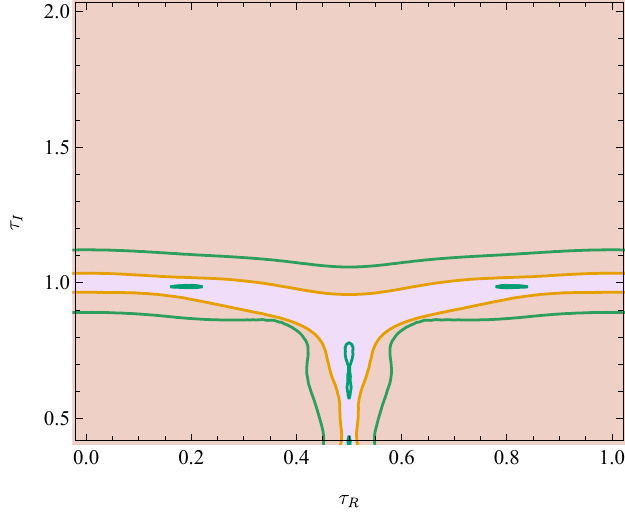}
	\caption{The parameter spaces of the scalar potential \eqref{eq:model2-potential} with $\beta^2 = 6.2972$. The scalar potential turns out to be steep.}
	\label{fig:model2-tausrpns}
\end{figure}

In this appendix, we discuss the other direction of the modification with the superpotential correction.
Instead of the introduction of the additional term in the K\"{a}hler potential~\eqref{eq:model1-deltaK}, let us consider the following additional matter contribution
\begin{align}
	W &= \Lambda^2 \bigl[
		Y^{3,6}_{\bm{1}} X + \beta Y^{3,4}_{\bm{1}} X' 
	\bigr],
	\label{eq:model2-superpotential}
	\\
	K &= - h \log ( -\iu \tau + \iu \bar{\tau}) + \frac{|X|^2}{(-\iu \tau + \iu \bar{\tau})^{-k_X}} + \frac{|X'|^2}{(- \iu \tau + \iu \bar{\tau})^{-k_{X'}}},
	\label{eq:model2-Kahler-potential}
\end{align}
where $k_{X'}$ denotes the modular weight of $X'$, and $\beta$ is a parameter associated with the additional contribution in the superpotential.
The scalar potential is given by 
\begin{align}
	V = ( 2 \tau_I)^6 |Y^{3,6}_{\bm{1}}|^2 + \beta^2 (2 \tau_I)^4 |Y^{3,4}_{\bm{1}}|^2,
	\label{eq:model2-potential}
\end{align}
where we have assumed $X \ll 1 $ and $X' \ll 1$.
As discussed in Sec.~\ref{sec:model}, the $\tau_I$ 
dependence in the scalar potential is determined by the modular weights of modular forms.
The profile of $(2 \tau_I)^6 |Y^{3,6}_{\bm{1}}|^2$ and $(2\tau_I)^4 |Y^{3,6}_{\bm{1}}|^2$ in $\tau_I = 1$ section is shown in the left panel of Fig.~\ref{fig:model2-potential}.
With a tuning of $\beta$, it seems possible to realize an apparent flat potential in $\tau_R$ direction \cite{Czerny:2014wza,Czerny:2014xja} at the first sight, because there is the relative phase shift of $\pi$
between two modular forms in the superpotential.
This can be seen from the $q$ expansion \eqref{eq:Y^{3,4}_1-q-exp} and \eqref{eq:Y^{3,6}_1-q-exp},
\begin{align}
	Y^{3,4}_{\bm{1}} \sim 1 + 240 q,
	\qquad
	Y^{3,6}_{\bm{1}} \sim 1 - 504 q,
\end{align}
where $q = \ee^{2 \pi \iu \tau}$.
The scalar potential with $\beta^2 = 6.2972$ is shown in the right panel of Fig.~\ref{fig:model2-potential} and 
there seems to exist a flat hilltop in the $\tau_R$ direction of the scalar potential \eqref{eq:model2-potential}.
However, $\tau_I$ in this potential is not stabilized around the hilltop and hence that direction still steep for realizing the successful slow-roll inflation as shown in Fig.~\ref{fig:model2-tausrpns}, where colored region shows the slow-roll parameters are bigger than unity. 
This is one of motivations to introduce $\delta K$ into the K\"{a}hler potential.

\section{Inflation rolling into other vacuum}
\label{app:other-vac-inflation}

\begin{figure}[t]
    \centering
    \includegraphics[width=0.48\textwidth]{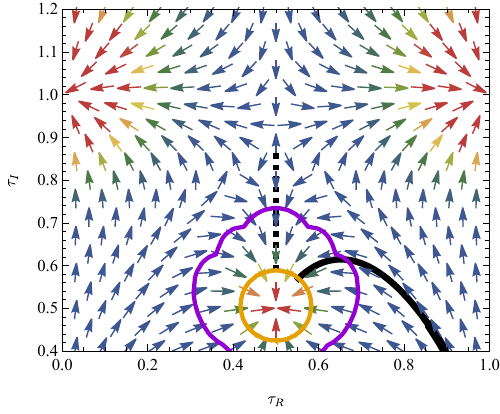}
    \quad
    \includegraphics[width=0.48\textwidth]{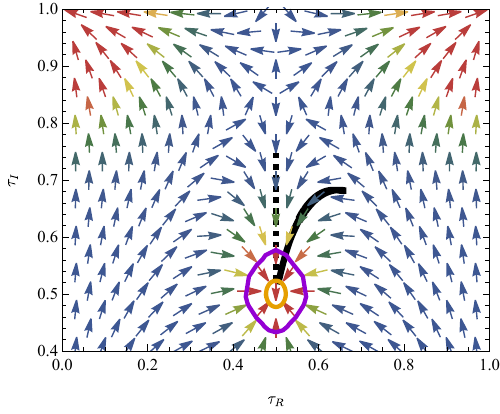}
    \caption{The vector plots of the potential gradients $- \frac{K^{ab} V_{,b}}{V}$ and inflationary trajectories going into the vacuum at $\tau = \frac{1 + \iu}{2}$.
    For the details, see the caption of Fig.~\ref{fig:model1-vectraj-alp}.}
    \label{fig:model1-vectraj-alp-vac2}
\end{figure}

\begin{figure}[t]
    \centering
    \begin{subfigure}{0.48\textwidth}
        \centering
        \includegraphics[width=\textwidth]{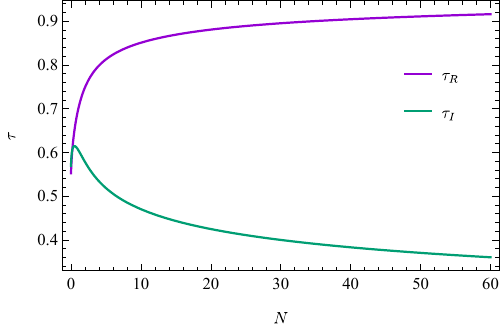}
        \\
        \includegraphics[width=\textwidth]{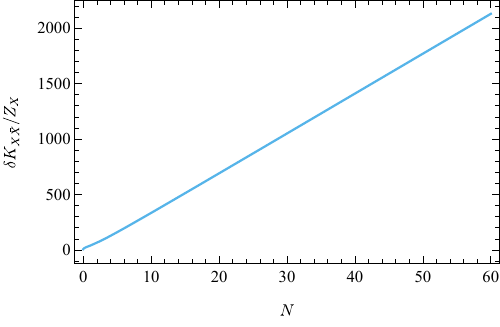}
        \caption{$\alpha = 0.1$}
    \end{subfigure}
    \quad 
    \begin{subfigure}{0.48\textwidth}
        \centering
        \includegraphics[width=\textwidth]{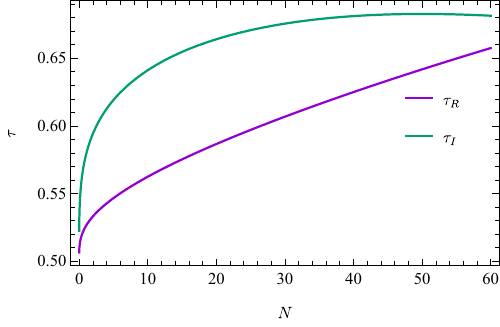}
        \\
        \includegraphics[width=\textwidth]{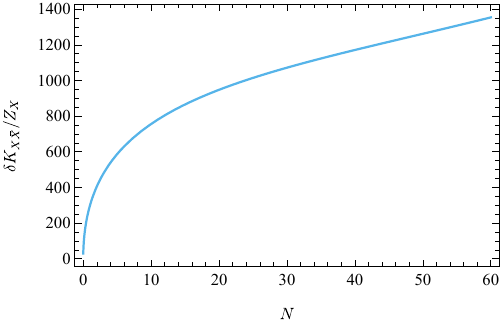}
        \caption{$\alpha = 5$}
    \end{subfigure}
    \caption{The time evolution of the moduli fields on the black solid lines in Fig.~\ref{fig:model1-vectraj-alp-vac2}. $\alpha$ is chosen as $0.1$ (left) and $5$ (right).
	}
	\label{fig:model1-Ntau-p1-vac2}
\end{figure}

\begin{figure}[ht]
	\centering
    \renewcommand\thesubfigure{\alph{subfigure}$^{\prime}$}
    \begin{subfigure}{0.48\textwidth}
        \centering
        \includegraphics[width=\textwidth]{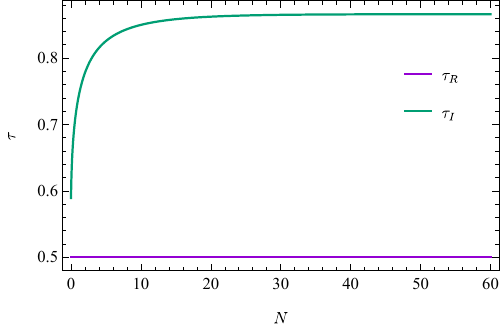}
        \\
        \includegraphics[width=\textwidth]{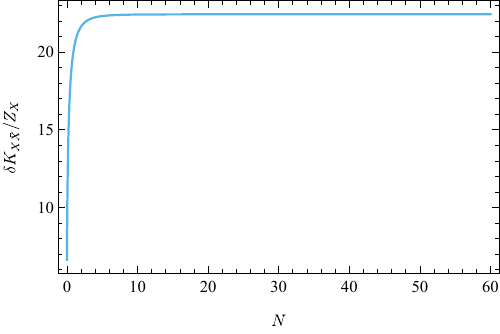}
        \caption{$\alpha = 0.1$}
    \end{subfigure}
    \quad 
    \begin{subfigure}{0.48\textwidth}
        \centering
        \includegraphics[width=\textwidth]{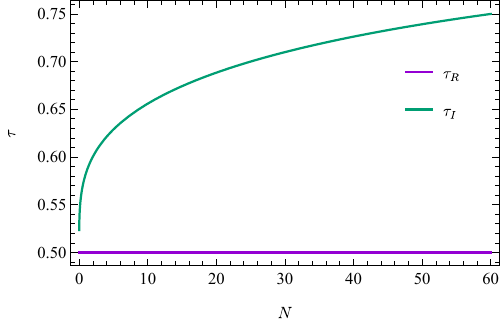}
        \\
        \includegraphics[width=\textwidth]{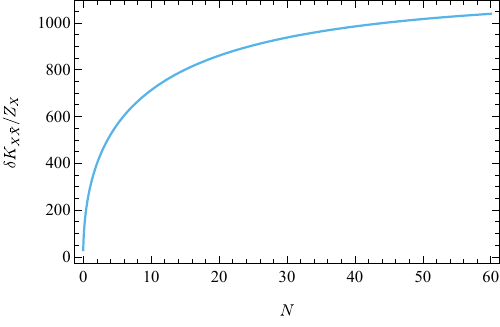}
        \caption{$\alpha = 5$}
    \end{subfigure}
    \caption{The time evolution of the moduli fields on the black dashed lines in Fig.~\ref{fig:model1-vectraj-alp-vac2}.
    }
    \label{fig:model1-Ntau-p2-vac2}
\end{figure}

In this section, we discuss the inflationary trajectories rolling into the vacuum at $\tau = \frac{1 + \iu}{2}$, which is identical to $\tau = \iu$ under the $S$ and $T$ modular transformations. The slow-roll inflation turns out to be similarly feasible around this vacuum as shown below.
As arrows seen in Fig.~\ref{fig:model1-vectraj-alp}, if the initial value of the modulus is $\tau_I \lesssim 1$, the moduli fields settle down into this vacuum after the inflation.
Fig.~\ref{fig:model1-vectraj-alp-vac2} shows such two trajectories starting at $N=60$ for $\alpha=0.1$ (left panel) and $5$ (right panel). Black solid lines show the inflationary trajectories where a combination of moduli including $\tau_R$ plays the inflaton, whereas black dashed ones show the similar trajectories where $\tau_I$ drives the inflation in terms of pole inflation.
Figs.~\ref{fig:model1-Ntau-p1-vac2} and \ref{fig:model1-Ntau-p2-vac2} show the time evolution of the moduli fields along the solid lines and dashed ones in Fig.~\ref{fig:model1-vectraj-alp-vac2}, respectively.
On the trajectories, the inflation can be realized but the perturbativity of $\delta K$ during the inflation is not obvious because the small $\tau_I$ makes $\delta K$ large.

\medskip
We have summarized in Tab.~\ref{tab:values-vac2} the spectral index $n_s$, tensor-to-scalar ratio $r$, power spectrum ${\cal P}_{\cal R}$, the overall scale $\Lambda$, and $|F_X|$ in the vacuum for each inflationary trajectory in Figs.~\ref{fig:model1-Ntau-p1-vac2} and \ref{fig:model1-Ntau-p2-vac2}.
It is found that the inflaton driven by $\tau_I$ 
for $\alpha =0.1$ produces too small $n_s$, which is inconsistent with the current observation.

\begin{table}[t]
    \centering
    \begin{tabular}{|c||c|c|c|c|c|}\hline
        trajectory & $n_s$ & $r$ & $\mc{P}_{\mc{R}} / (\Lambda /M_P)^4$ & $\Lambda/ M_P$ & $|F_X/M_P^2|$ \\
        \hhline{|=#=|=|=|=|=|}
        (a) in Fig.~\ref{fig:model1-Ntau-p1-vac2}& $0.966$ & $6.33 \times 10^{-5}$ & $1.07 \times 10^4$ & $6.66 \times 10^{-4}$ & $2.23 \times 10^{-10}$ \\
        \hline
        (a$^\prime$) in Fig.~\ref{fig:model1-Ntau-p2-vac2}& $0.744$ & $7.69 \times 10^{-11}$ & $8.41 \times 10^9$ & $2.24 \times 10^{-5}$ & $2.52 \times 10^{-13}$\\
        \hline
        (b) in Fig.~\ref{fig:model1-Ntau-p1-vac2}& $0.993$ & $4.12 \times 10^{-5}$ & $3.28 \times 10^2$ & $1.60 \times 10^{-3}$ & $1.27 \times 10^{-9}$ \\
        \hline
        (b$^\prime$) in Fig.~\ref{fig:model1-Ntau-p2-vac2}& $0.965$ & $1.40 \times 10^{-5}$ & $9.67 \times 10^2$ & $1.21 \times 10^{-3}$ & $7.42 \times 10^{-10}$ \\
        \hline
    \end{tabular}
    \caption{Values of $n_s$, $r$, ${\cal P}_{\cal R}$, $\Lambda$ in the superpotential, and $F_X$ for each trajectory in Figs.~\ref{fig:model1-Ntau-p1-vac2} and \ref{fig:model1-Ntau-p2-vac2}.
    $n_s$ and $r$ are evaluated at $N=60$.}
    \label{tab:values-vac2}
\end{table}

\newcommand{\arxivfont}{\rmfamily}
{\small
\bibliographystyle{yautphys}
\bibliography{ref-mfinflation}
}

\end{document}